\documentclass[aps,prr,superscriptaddress,twocolumn]{revtex4-2}
\usepackage{bm}
\usepackage{amsmath}
\usepackage{amsfonts}
\usepackage{amssymb}
\usepackage{graphicx}
\usepackage{color}
\usepackage{xcolor}
\usepackage[colorlinks,linkcolor=blue,anchorcolor=blue,citecolor=blue,urlcolor=blue]{hyperref}
\usepackage{dcolumn}
\usepackage{fancyhdr}
\usepackage[normalem]{ulem}

\begin{document}

\title{Quantum phases in spin-orbit-coupled Floquet spinor Bose gases}
	
\author{Yani Zhang}
\affiliation{Department of Physics, Shanghai University, Shanghai 200444, China}

\author{Yuanyuan Chen}
\email{cyyuan@shu.edu.cn}
\affiliation{Department of Physics, Shanghai University, Shanghai 200444, China}

\author{Hao Lyu}
\email{lyuhao@shu.edu.cn}
 \affiliation{Quantum Systems Unit, Okinawa Institute of Science and Technology Graduate University, Onna, Okinawa 904-0495, Japan}	
	
\author{Yongping Zhang}
\email{yongping11@t.shu.edu.cn}
\affiliation{Department of Physics, Shanghai University, Shanghai 200444, China}

\begin{abstract}

We propose a spin-orbit-coupled Floquet spinor Bose-Einstein condensate (BEC) which can be implemented by Floquet engineering of a quadratic Zeeman field. The Floquet spinor BEC has a Bessel-function-modulated Rabi frequency and a Floquet-induced spin-exchange interaction.  
The quantum phase diagram of the spin-orbit-coupled Floquet spinor BEC is investigated by considering antiferromagnetic or ferromagnetic spin-spin interactions. 
In comparison with the usual spin-orbit-coupled spin-1 BEC, 
we find that a stripe phase for antiferromagnetic interactions  can exist in a large quadratic Zeeman field regime, 
and a different stripe phase with an experimentally favorable contrast for ferromagnetic interactions is uncovered.

\end{abstract}
	
\maketitle

\section{introduction}
\label{introduction}

Ultracold neutral atoms provide a fertile  playground for  engineering artificial gauge fields~\cite{Goldman,Galitski,Zhai2015,Zhang2016}.
Synthetic spin-orbit coupling, utilizing atomic hyperfine levels as pseudo-spins, can be realized by coupling these states via Raman lasers~\cite{Lin,WangP,Cheuk}.
Spin-orbit-coupled Bose-Einstein condensates (BECs) open a new route to explore exotic superfluid states and simulate topological matter~\cite{JiS,WuZ,HuangL,Mossman,Valdes,Frolian}.  
One interesting feature is that the spin-orbit coupling modifies the dispersion relation of a BEC.  
The spin-orbit-coupled dispersion may have multiple energy minima. 
Condensations into these energy minima present exotic quantum phases, such as the plane-wave (PW) phase and stripe phase
~\cite{Wang2010,Wucong2011,Hotian2011,Hu2012,Yongping2012,LiY2012}.
The PW phase occupies one of the minima and possesses a nonzero quasimomentum, which breaks the time-reversal symmetry~\cite{Wucong2011}. 
The phase transition and excitations of  PW states have been experimentally observed~\cite{JiS,Khamehchi}.  
The stripe phase, condensing at least two minima, represents a combination of spatial density modulation and superfluidity and is identified as having supersolid properties~\cite{LiY2013}.
The realization of the stripe phase requires miscibility of the two spin components  and a low Rabi frequency of the Raman lasers~\cite{LiY2012, Zheng2013}.
This is quite a challenge in $^{87}$Rb atoms experiments since atomic interactions are insensitive to the hyperfine states~\cite{Martone2014, Luo2019,Peter2019}. 
Recently, the spin-orbit-coupling-induced stripe phase was observed in atoms loaded into superlattices~\cite{LiJR}, 
in which the sub-lattice sites are treated as pseudo-spins.

A spinor BEC has more degrees of freedom and intriguing interactions which lead to a rich ground-state phase diagram~\cite{Stamper-Kurn}. 
A spin-orbit-coupled spin-1 BEC has been experimentally realized~\cite{Campbell}. 
Quantum phases in spin-orbit-coupled spin-1 BECs depend on antiferromagnetic and ferromagnetic spin-spin interactions and show salient features~\cite{Lan,Sun, Yu,Martone2016,ChenY}. 
Three different kinds of stripe phases have been revealed to exist, and phase transitions between different phases are so abundant that tricritical  points emerge~\cite{Yu,Martone2016}. 
One of outstanding features is that the quadratic Zeeman field plays an important role in the existence of stripe phases.  
Especially, in a ferromagnetic spinor BEC, stripes appear in the limited regime of low Rabi frequency and quadratic Zeeman field~\cite{Campbell}.  

On the other hand, Floquet engineering is a powerful tool in quantum physics for controlling system parameters and manipulating quantum states~\cite{Bukov,Eckardt,Oka}.
In a periodically driven system, an effective static Hamiltonian can be tailored which depends on the driving parameters.
The driving could lead to dramatic changes in the system properties.
Ultracold atoms provide an ideal platform for Floquet engineering due to the tunability and purity of the system,
which has been used to explore artificial gauge fields, topological insulators, and soliton dynamics~\cite{Jotzu,Struck,GoldmanPRX,Flaschner,Ha,Schweizer,Wintersperger,Mitchell,LuM}. 
In spin-orbit-coupled ultracold atoms, a coherently periodic modulation of Raman laser intensities can produce a tunable spin-orbit coupling strength~\cite{Zhang2013,Jimenez-Garcia,Llorente}, which provides a practical way for dynamical control. 
A periodic modulation of Raman laser frequencies is employed to manipulate the emergence of the Dirac point in Floquet bands and thus to change band topology~\cite{Huang2018}.  
A shaking Raman lattice that generates high-dimensional spin-orbit coupling is implemented to tune Floquet topological bands~\cite{ZhangJY}.
Recently, a Floquet spinor BEC was proposed using a periodically driven quadratic Zeeman field~\cite{Kazuya}. 
Compared with the usual spinor BEC, the Floquet spinor BEC has an additional spin-exchange interaction which is induced by the high-frequency driving. 
Such an induced spin-exchange interaction can have a profound effect in ferromagnetic condensates and can generate unconventional quantum phases~\cite{Kazuya}. 

In this paper, we study a Floquet spin-1 BEC with spin-orbit coupling. In spin-1 spin-orbit coupling experiments, three external Raman lasers are used to couple three hyperfine states, 
and the quadratic Zeeman effect is proportional to the two-photon detunings between Raman lasers and hyperfine states~\cite{Campbell}.  
We propose to drive the quadratic Zeeman effect periodically around a constant value via periodic modulation of Raman laser frequencies.   
Under high-frequency driving, the spin-orbit-coupled spinor BEC is effectively described by a static Floquet spinor BEC, in which the Rabi coupling is modulated by a Bessel function and a unique spin-exchange interaction emerges. Quantum ground phases are investigated in such a spin-orbit-coupled Floquet spinor BEC with antiferromagnetic or ferromagnetic spin-spin interactions. Our main results are as follows.

(i) Due to the Bessel-function-modulated Rabi frequency, each quantum phase can exist in a broad region of the Rabi frequency.  
Previous studies showed that the stripe phases in antiferromagnetic and ferromagnetic spinor BECs exist in a small regime of the Rabi frequency, 
say, $\Omega_{c1} <\Omega <\Omega_{c2} $, where $\Omega$ is the Rabi frequency and $\Omega_{c1,c2}$ are two critical values, 
with $\Omega_{c2}-\Omega_{c1}$ being a small quantity~\cite{Campbell,Yu,Martone2016}.  
In the Floquet spinor BEC, the Rabi frequency  is modulated as $\Omega J_0$ with $J_0$ being the zero-order Bessel function of the first kind.  
We find that the corresponding phases appear in $\Omega_{c1}/J_0 <\Omega <\Omega_{c2}/J_0 $. 
Since $J_0$ is tunable and less than 1, the $\Omega$ region for the existence of the stripe phase is enlarged significantly. 
This extension of the Rabi frequency for the stripe phases is beneficial for their experimental observations. 

(ii) For antiferromagnetic interactions, the appearance of the Floquet-induced spin-exchange interaction extends the second stripe phase to broaden the quadratic Zeeman field domain, which exists in an extremely narrow region of the quadratic Zeeman field in a typical spin-orbit-coupled spinor BEC. 

(iii) For ferromagnetic interactions, a different stripe phase is induced by the combined effects of the Floquet-induced spin-exchange interaction and the Rabi coupling.  These stripes have a very high density contrast.  Their Bogoliubov excitations are identified as having two gapless Nambu-Goldstone modes.

This paper is organized as follows. 
In Sec.~\ref{model}, we present the theoretical model for a spin-orbit-coupled Floquet spinor BEC. 
It features the Floquet-induced spin-exchange interaction and the Bessel-function-modulated Rabi frequency. 
In Sec.~\ref{noninteracting}, the phase diagram of a noninteracting spin-orbit-coupled Floquet spinor BEC is analyzed. 
In Sec.~\ref{interacting}, phase diagrams for antiferromagnetic and ferromagnetic spin-spin interactions are studied separately. 
Finally, the conclusion follows in Sec.~\ref{conclusion}.

\section{MODEL}
\label{model}

We consider an experimentally realizable  spin-orbit-coupled spin-1 BEC. The spin-orbit coupling is implemented by coupling three hyperfine states with total angular momentum $F=1$ ($m_F=0,\pm1$)
via three Raman lasers propagating along the $x$ axis~\cite{Campbell}. 
Adjusting two-photon detunings between Raman lasers and hyperfine states so that they are equal can mimic an effective quadratic Zeeman field. 
We propose to periodically drive it by a periodic oscillation of the Raman laser frequencies. The mean-field energy functional of the oscillating system is
\begin{align}
E[\Phi]&=\int d\bm{r} \Phi^\dagger \left[ H_{\text{SOC}}+ \xi(t)  F_z^2 \right] \Phi \notag\\
&\phantom{={}}+\int d\bm{r} \Phi^\dagger\left[\frac{c_0}{2} \Phi^\dagger \Phi+\frac{c_2}{2} 
\Phi^\dagger \bm{F} \Phi\cdot \bm{F}  \right] \Phi,
\label{EnergyS}
\end{align}
with $\Phi=(\Phi_1,\Phi_2,\Phi_3)$ being the spin-1 spinor describing three-component wave functions.
$\bm{F}=(F_x,F_y,F_z)$ are the spin-1 Pauli matrices.  $H_{\text{SOC}}$ is the single-particle spin-orbit-coupled Hamiltonian, 
\begin{align}
H_{\text{SOC}}=\left(-i\frac{\partial}{\partial x} +2F_z\right)^2 + \varepsilon F^2_{z} +\frac{\Omega}{\sqrt{2}} F_{x},
\label{SocH}
\end{align}
where $\Omega$ is the Rabi frequency depending on the laser intensities and $\varepsilon$ is a constant quadratic Zeeman shift which is induced
by the detunings of the Raman lasers~\cite{Campbell}. 
The spin-1 spin-orbit coupling is represented by the second term in Eq.~(\ref{SocH}) with a fixed coupling strength due to the experimentally chosen gauge.
In our dimensionless equations, the units of momentum, length, and energy are $\hbar k_{\text{Ram}}$, $1/k_{\text{Ram}}$, 
and $E_\text{R}=\hbar^2k^2_{\text{Ram}}/2m$, respectively.
Here, $m$ is the atom mass, and $k_{\text{Ram}} = 2\pi/\lambda_{\text{Ram}}$ is the wave number of the Raman lasers, 
with $\lambda_{\text{Ram}}$ being the wavelength. 
Considering the typically experimental parameter $\lambda_{\text{Ram}}=790$ nm, 
we have $E_\text{R}=2\pi \hbar \times  3.67$ kHz as the units of energy for rubidium atoms~\cite{Campbell}. 
The quadratic Zeeman field is periodically driven,
\begin{equation}
\xi(t)=\alpha \cos(\omega t),
\end{equation}
with $\omega$ and $\alpha$ being
the frequency and amplitude of  the driving, respectively.
$c_{0}$ and $c_{2}$ in Eq.~(\ref{EnergyS}) denote density-density and spin-spin interactions, respectively, 
which depend on the $s$-wave scattering lengths in the total spin channels. 
In this work, we assume a repulsive density-density interaction ($c_0>0$), while the 
spin-spin interaction $c_2$ can be either positive (antiferromagnetic) or negative (ferromagnetic).

For high-frequency driving, we can derive an effective static Hamiltonian by averaging the time-dependent Hamiltonian over one modulation period~\cite{Eckardt}.
We transform the system into an oscillating  frame by using the transformation,
\begin{align}
U(t)=\exp\left(-i\frac{\alpha}{\omega}\sin(\omega t)F_z^2\right).
\end{align}
After applying the transformation $\Phi=U(t)\Psi$, the resultant time oscillating terms are dropped due to the average in a period. 
Finally, we end up with the following time-independent energy functional:
\begin{align}
E[\Psi]&=\int d\bm{r} \Psi^\dagger\left[ H^{\prime}_{\text{SOC}}
+\frac{c_0}{2} \Psi^\dagger \Psi+\frac{c_2}{2} 
\Psi^\dagger \bm{F} \Psi\cdot \bm{F}  \right]  \Psi \notag\\
&\phantom{={}} +c_{f}\int d\bm{r} \left( \Psi_{1}^{\dagger}\Psi_{3}^{\dagger}\Psi^{2}_{2}+ \Psi_{1}\Psi_{3}\Psi_{2}^{\dagger2}\right).
\label{eq:energy}
\end{align}
The energy functional describes a spin-orbit-coupled Floquet spinor BEC with the spinor $\Psi=(\Psi_1,\Psi_2,\Psi_3)$.  
The modulated single-particle Hamiltonian is 
\begin{align}
H^{\prime}_{\text{SOC}}&=\left(-i\frac{\partial}{\partial x} +2F_z\right)^2 + \varepsilon F^2_{z} 
+\frac{\Omega}{\sqrt{2}}J_0\left(\frac{\alpha}{\omega}\right) F_{x}.
\label{eq:SOC}
\end{align}
Note that the only difference between the Floquet spin-orbit coupled Hamiltonian $H^{\prime}_{\text{SOC}}$ and the original one $H_{\text{SOC}}$ is that the Rabi frequency is modulated by the zero-order Bessel function of the first kind $J_0(\alpha/\omega)$. 
The density-density and spin-spin interactions in Eq.~(\ref{eq:energy}) are the same as those in the usual spinor BEC. 
Nevertheless, there is a new spin-exchange interaction with the coefficient $c_f$ which is a pure effect of Floquet modulation~\cite{Kazuya},
\begin{equation}
c_{f}=c_{2}\left[1-  J_{0}\left(2 \alpha /\omega \right)\right]. 
\end{equation}
The spin-orbit-coupled Floquet spinor BEC is reduced back to the usual spin-orbit-coupled spinor BEC if the driving disappears, i.e., $\alpha/\omega=0$.

\section{Phase diagram of the noninteracting spin-orbit-coupled Floquet spinor BEC}
\label{noninteracting}

We study quantum phases in a spin-orbit-coupled Floquet spinor BEC.  
First, we analyze the single-particle phase diagram, which was addressed in Refs.~\cite{Lan,Sun, Yu}. 
An analysis of the single-particle phase diagram can provide insight into ground states in an interacting system. 
The dispersion of $H^{\prime}_{\text{SOC}}$  can be calculated by direct diagonalization. 
Depending on spin-orbit coupling parameters, the lowest band in the dispersion may have one, two or three local minima.  
Ground states choose one of  minima to occupy. Therefore, a general ground-state wave function should be
\begin{align}
\Psi=\sqrt{\bar{n}}e^{ikx}\left(\begin{array}{c}
\cos\theta\cos\varphi  \\  -\sin\theta \\ \cos\theta\sin\varphi
\end{array}\right),
\label{eq:pw}
\end{align}
where $\bar{n}=N/V$, with $N$ being the total atom number and $V$ the volume of the system;
$k$ is the quasimomentum; and $\theta$ and $\varphi$ are two parameters.  
By substituting Eq.~(\ref{eq:pw}) into Eq.~(\ref{eq:energy}) (with $c_0=c_2=0$), we obtain the energy per particle,
 \begin{align} 
E_{k}=k^{2}- \left( \frac{A_{k}^{\prime}}{54}\right)^{\frac{3}{2}}-A_{k} \left(\frac{2} {27 A_{k}^{\prime}}\right)^{\frac{3}{2}}  +\frac{2}{3} A_{0},
\end{align}
with 
\begin{align}
A_{k}&=48 k^{2}+(\varepsilon+4)^2+\frac{3}{2}J^2_{0}\left(\frac{\alpha}{\omega}\right) \Omega^{2},\notag\\
A^{\prime}_{k}&=(\varepsilon+4)A_{k}^{\prime \prime}+\sqrt{(\varepsilon+4)^{2} A_{k}^{\prime \prime 2}-4 A_{k}^{3}} ,  \notag\\
A_{k}^{\prime \prime}&=-288 k^{2}+2(\varepsilon+4)^{2}+\frac{9}{2}J^2_{0}\left(\frac{\alpha}{\omega}\right) \Omega^{2}. \notag
\end{align}
Then the quasimomentum can be determined by solving $\partial E_k/\partial k=0$. The occupation of $k=0$ is the zero-momentum (ZM) state, and the occupation of a nonzero quasimomentum is the PW state.

%%%%%%%%%%%%%%%%%%%%%%%%%%%%%%%%%%%%%%%%%%%%%%%%%%%%%%%%
\begin{figure}[t]
\includegraphics[width=3.2in]{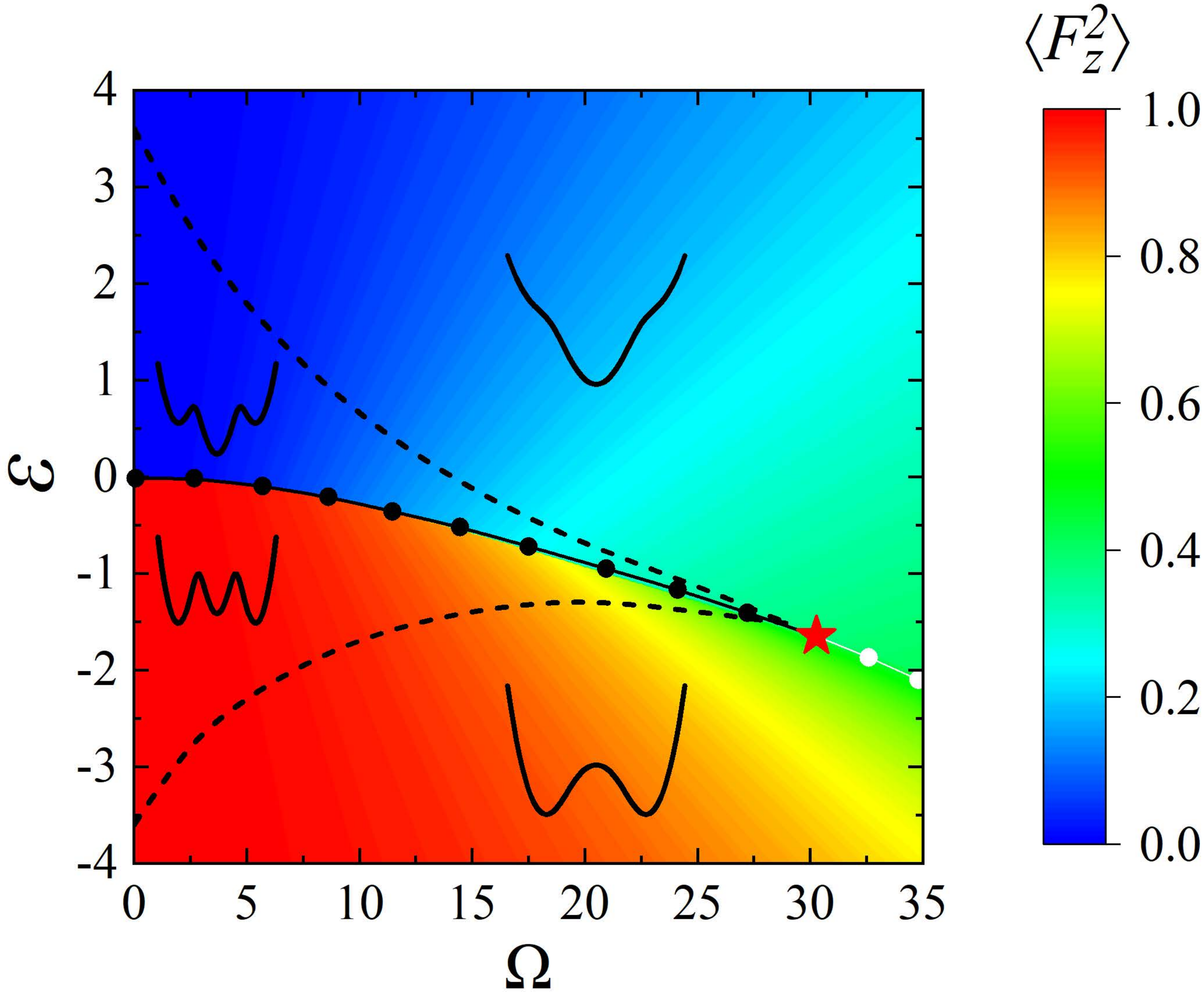}
\caption{Quantum ground-state phase diagram of a noninteracting spin-orbit-coupled  Floquet spinor BEC in the space of the Rabi frequency $\Omega$ and the quadratic Zeeman field $\varepsilon$. 
The driving is $\alpha/\omega=2$ [$J_0(\alpha/\omega)=0.224$].  
The background corresponds to values of the tensor magnetization $\langle {F}^2_z \rangle$. 
The black and white solid lines with dots represent first-order and second-order phase transitions, respectively. 
Below these lines is the plane-wave phase, and beyond is the zero-momentum phase. 
The red star denotes a tricritical point. Insets show the lowest bands of the single-particle dispersion. 
The black dashed lines separate different regions where the lowest band of the dispersion has one, two or three local energy minima. 
}
\label{Fig1}
\end{figure}
%%%%%%%%%%%%%%%%%%%%%%%%%%%%%%%%%%%%%%%%%%%%%%%%%%%%%%%%

Figure~\ref{Fig1} shows the ground-state phase diagram in the $(\Omega, \varepsilon)$ plane,  
in which  the tensor magnetization $\langle {F}^2_z \rangle=\cos^2\theta$ is chosen as the order parameter. 
The solid lines with dots are the transition lines between PW and ZM phases, above which is the ZM phase and below which is the PW phase.  
We also show the lowest band of $H^{\prime}_{\text{SOC}}$ in Fig.~\ref{Fig1}. 
The dashed line in the ZM regime is a separation, above which the lowest band has only one minimum at $k=0$ and below which it has three local minima but the lowest one at $k=0$.  
In the PW regime, the lowest band may have two or three local minima. 
The separation between these two cases is demonstrated by the black dashed lines. 
Two dashed lines merge together with the phase transition line at the so-called tricritical point, which is labeled by the red star in Fig.~\ref{Fig1}. 
The location of the tricritical point can be analytically calculated from $\partial^2 E_k/\partial k^2=0$ and the equal energy between the PW and ZM states~\cite{Sun,Yu}. The calculated value for the tricritical point is $(\Omega^{\ast},\varepsilon^{\ast})=(30.14,-1.66)$.  
When  $\Omega<\Omega^{\ast}$ the PW-ZM transition is first-order and when $\Omega>\Omega^{\ast}$ the phase transition is second-order.

\section{Phase diagram of the interacting spin-orbit-coupled Floquet spinor BEC}
\label{interacting}

For a spin-orbit-coupled spin-1 BEC, previous works revealed ground states, including PW, ZM, and stripe phases and rich phase transitions between them~\cite{Campbell, Sun,Yu,Martone2016}. The single-particle dispersion of spin-orbit coupling provides diverse arrangements of energy minima, and interactions determine how they condense into these minima.   
Since the dispersion of  $H^{\prime}_{\text{SOC}}$ has three energy-minima at most, 
we construct ground-state wave functions as a superposition of the spinors at these minima, which can be assumed to be
\begin{align}
\Psi&=\sqrt{\bar{n}} C_{+}e^{ikx}\left(\begin{array}{c}
\cos\theta_1\cos\varphi  \\  -\sin\theta_1 \\ \cos\theta_1\sin\varphi
\end{array}\right)+\sqrt{\bar{n}} C_{0}\left(\begin{array}{l}
\sin\theta_2/\sqrt{2}  \\  -\cos\theta_2 \\ \sin\theta_2/\sqrt{2}
\end{array}\right) \notag\\
&\phantom{={}}+\sqrt{\bar{n}}C_{-}e^{-ikx} \left(\begin{array}{c}
\cos\theta_1\sin\varphi \\ -\sin\theta_1 \\ \cos\theta_1\cos\varphi
\end{array}\right).
\label{eq:variation}
\end{align}
The superposition coefficients satisfy the normalization condition, $|C_{+}|^2+|C_0|^2+|C_{-}|^2=1$.  
The spinors are the eigenstates of $H^{\prime}_{\text{SOC}}$, with the exact parameters $\theta_{1,2}$ and $\varphi$ to be specified.  
The second state in Eq.~(\ref{eq:variation}) is the spinor at $k=0$,  and the first and third ones are spinors modulated by the plane waves at $\pm k$. 
The symmetry of $H^{\prime}_{\text{SOC}}$ requires that the first and third states have the same $\theta_1$ and $\varphi$. We substitute the above variational wave functions (\ref{eq:variation}) into the energy functional in Eq.~(\ref{eq:energy}). 
The minimization of the resultant energy functional gives the values of parameters $k$, $C_{0,\pm}$, $\theta_{1,2}$, and $\varphi$. 
From these parameters, we can classify ground states: 
the ZM phase has $C_\pm=0$; the PW phase has a nonzero $k$ and $C_0=0$, with one of  $C_\pm$ being nonzero; 
and the stripe phase requires $k\ne 0$ and at least two of $C_{\pm,0}$ are nonzero. 
The stripe phases can be further classified according to relative values of $C_{\pm,0}$~\cite{Yu,Martone2016}. 
Considering that the classification of ground states depends strongly on $C_{\pm,0}$, we use the tensor magnetization $\left\langle F^2_z\right\rangle$ as the order parameter to identify different phases, 
\begin{equation}
\left\langle F^2_z\right\rangle=\left(\left|C_{+}\right|^2+\left|C_{-}\right|^2\right) \cos ^2 \theta_1+\left|C_0\right|^2 \sin ^2 \theta_2.
\label{order}
\end{equation}
We find that antiferromagnetic and ferromagnetic spin-spin interactions have very different ground-state phase diagrams,
which are studied separately.

\subsection{Antiferromagnetic interactions}

%%%%%%%%%%%%%%%%%%%%%%%%%%%%%%%%%%%%%%%%%%%%%%%%%%%%%%%%
\begin{figure}[t]
\includegraphics[ width=3.2in]{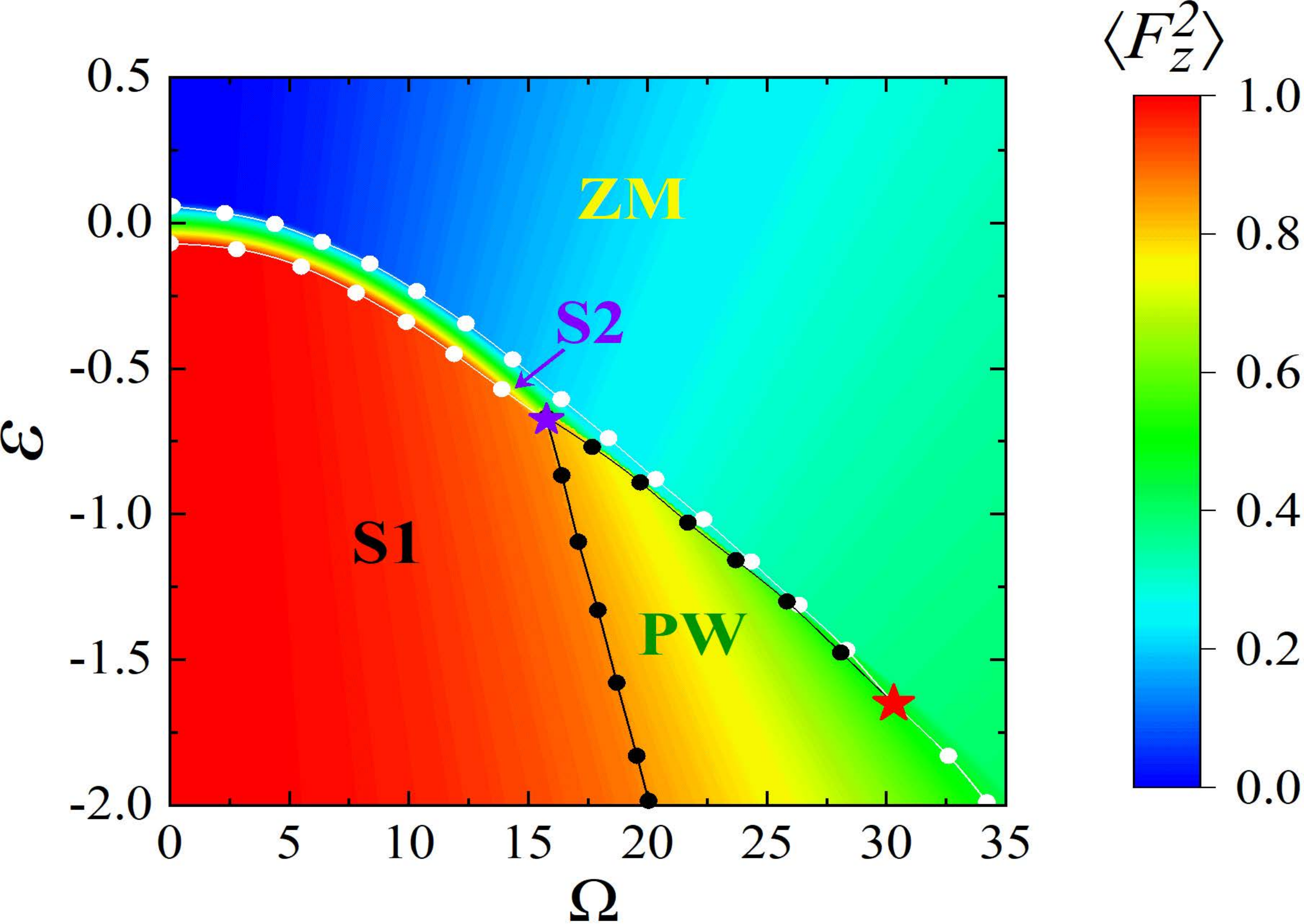}
\caption{Quantum ground-state phase diagram of a spin-orbit coupled Floquet spinor BEC with an antiferromagnetic spin-spin interaction ($\bar{n}c_0=1$ and $\bar{n}c_2=0.1$). 
The background corresponds to values of the tensor magnetization $\langle {F}^2_z \rangle$ defined in Eq.~(\ref{order}). 
The black and white solid lines with dots represent the first-order and second-order phase transitions, respectively.
The different tricritical points are denoted by the red and purple stars.
The driving is $\alpha/\omega=2$ [$J_0(\alpha/\omega)=0.224$ and $J_0(2\alpha/\omega)=-0.397$]. 
}
\label{Fig2}
\end{figure}
%%%%%%%%%%%%%%%%%%%%%%%%%%%%%%%%%%%%%%%%%%%%%%%%%%%%%%%%

The antiferromagnetic interaction is $c_2>0$, which is typical for the $^{23}$Na BEC.  
Figure~\ref{Fig2} demonstrates the phase diagram for antiferromagnetic interactions with driving $\alpha/\omega=2$ in the space of the quadratic Zeeman field $\varepsilon$ and the Rabi frequency $\Omega$.  
When $\varepsilon$ is negative, the single-particle dispersion has two lowest minima located at $\pm k_m$ (see the inset in Fig.~\ref{Fig1}); 
the antiferromagnetic interaction allows atoms to simultaneously occupy these two minima to form a stripe for low $\Omega$.   
This stripe phase labeled as S1 in Fig.~\ref{Fig2}, has $|C_{+}|=|C_{-}|=1/\sqrt{2}$ and $C_0=0$. 
Using the wave functions in Eq.~(\ref{eq:variation}) with  $C_0=0$ and considering the single-particle spinors at $\pm k_m$ with $\varphi= \pi/2$, 
we get the energy of the antiferromagnetic interaction $\langle E \rangle_{c_2}$ and Floquet-induced spin-exchange interaction $\langle E \rangle_{c_f}$,
\begin{align}
\langle E \rangle_{c_2}  +\langle E \rangle_{c_f}  &=\frac{c_2\bar{n}^2 }{2}\cos^4\theta_1  +  c_2\bar{n}^2|C_{-}|^2|C_{+}|^2
\notag \\  
&\phantom{={}}\times\left[(1+\frac{c_f}{c_2}) \sin^2(2\theta_1)-2\cos^4\theta_1\right].  
\label{Twoenergy}    
\end{align} 
For a low $\Omega$, we have $\theta_1\approx 0$, and the minimization of $\langle E \rangle_{c_2} +\langle E \rangle_{c_f}$ leads to $|C_{+}|=|C_{-}|=1/\sqrt{2}$, 
corresponding to the S1 phase, the tensor magnetization of which is $\left\langle F^2_z\right\rangle \approx 1$, as shown in  Fig.~\ref{Fig2}.  
$\theta_1$ prefers to be nonzero for a large $\Omega$.
Meanwhile, the first term 
$c_2\bar{n}^2/2\cos^4\theta_1$ in $\langle E \rangle_{c_2} +\langle E \rangle_{c_f}$ allows $\theta_1$ to approach  to $\pi/2$, 
at which it is minimized, so that  $\theta_1$ can grow from zero to $\pi/2$ as $\Omega$ increases. 
Consequently, for a high $\Omega$, we may have $(1+c_f/c_2) \sin^2(2\theta_1)-2\cos^4\theta_1>0$. Then the minimization of $\langle E \rangle_{c_2} +\langle E \rangle_{c_f}$ requires one of  $C_{\pm}$ to be zero. 
Even though the single-particle dispersion has two minima, the antiferromagnetic interaction chooses one of them to occupy, 
generating the PW phase shown in Fig.~\ref{Fig2}. 
The phase transition between the S1 and PW phases is first order. 
Physically,  $\langle E \rangle_{c_2} +\langle E \rangle_{c_f}$ is proportional to $c_2\bar{n}^2 |C_{+}|^2|C_{-}|^2 [(1+c_f/c_2)\langle F_x \rangle_+ \langle F_x \rangle_- + \langle F_z \rangle_+ \langle F_z \rangle_-]$, 
where $\langle F_x \rangle_\pm $ and $\langle F_z \rangle_\pm $ are the $x$ and $z$ polarizations of the spinors at $\pm k_m$. 
The antiferromagnetic interaction generates $ \langle F_z \rangle_+ \langle F_z \rangle_- <0$, and the Rabi coupling favors 
$\langle F_x \rangle_+ \langle F_x \rangle_->0$. 
The competition between these two effects gives rise to the S1-PW transition, and we have $\left\langle F^2_z\right\rangle < 1$ in the PW phase (see Fig.~\ref{Fig2}).
The emergence of the ZM phase in Fig.~\ref{Fig2} is due to the fact that the lowest minimum of the single-particle dispersion lays at $k=0$. 
There is a second stripe phase labeled S2 which is unique only for antiferromagnetic interactions. 
The S2 phase is featured with $|C_-|=|C_+|\ne 0,  |C_0| \ne 0$, and $\Theta\equiv \arg(C_-)+ \arg(C_+)-2\arg(C_0)=\pi$.

At first glance, the phase diagram shown in Fig.~\ref{Fig2} is similar to that of the usual spin-orbit-coupled BEC demonstrated in Refs.~\cite{Yu,Martone2016} (i.e., Fig.~1(a) in~\cite{Yu} and Fig.~1 in~\cite{Martone2016}).  
There are two tricritical points represented by stars in Fig.~\ref{Fig2}. The first- (second-) order phase transitions between different phases are shown by black (white) solid lines with dots.
However, there are two different features in our system.  
(i) All phases exist in a broadened region of the Rabi frequency. This is a straightforward consequence of the Bessel-function modulation $\Omega J_0$. 
(ii) The existence of the S2 phase is also extended in the $\varepsilon$ domain. 
In the usual spin-orbit-coupled antiferromagnetic BEC the S2 phase exists in an extremely narrow region of $\varepsilon$ (see Fig.~1(a) in~\cite{Yu} and Fig.~1 in~\cite{Martone2016}).  
Our Floquet system has a large extension, which benefits from the Floquet-induced interaction.

%%%%%%%%%%%%%%%%%%%%%%%%%%%%%%%%%%%%%%%%%%%%%%%%%%%%%%%%
\begin{figure}[t]
\includegraphics[ width=3.2in]{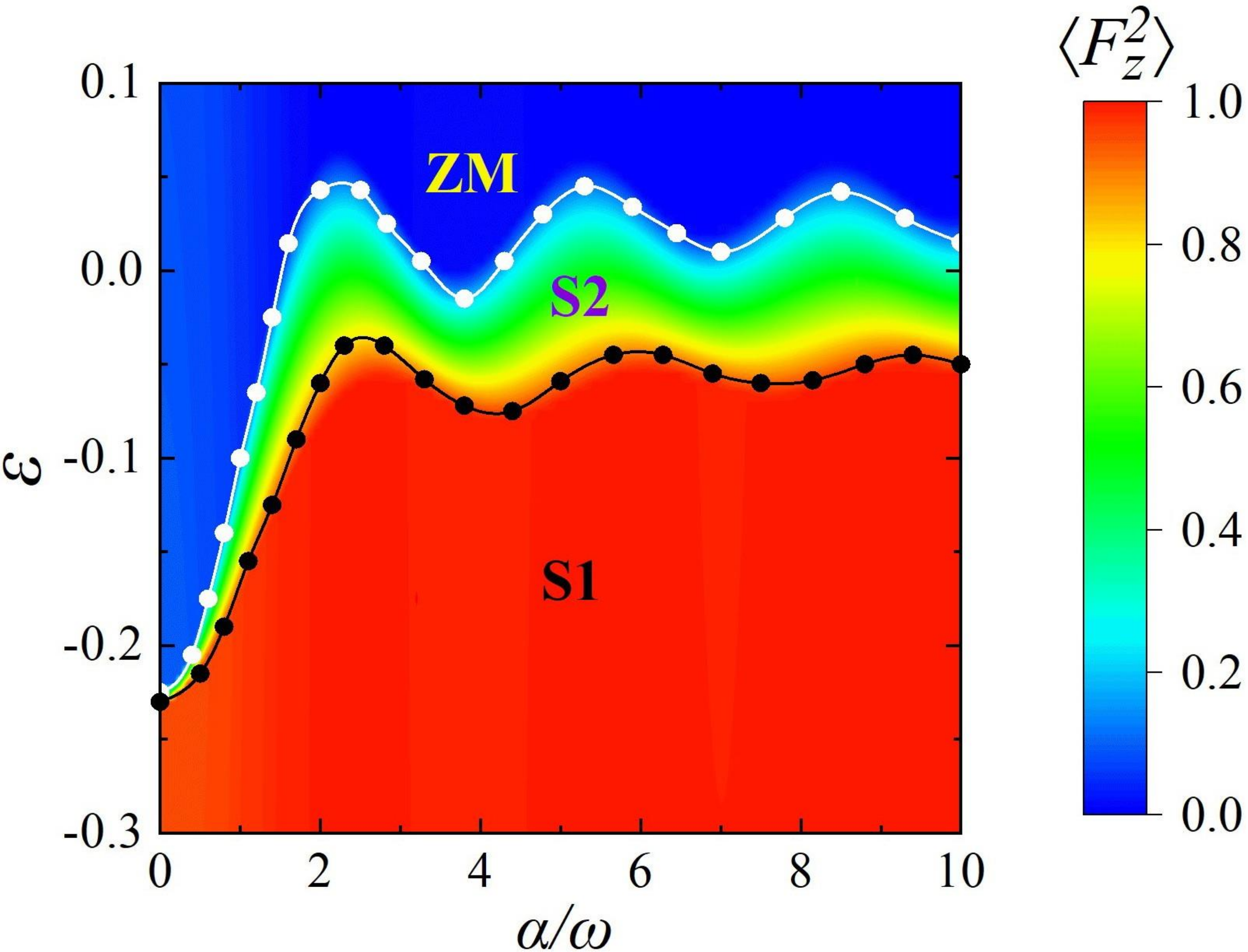}
\caption{Phase diagram for an antiferromagnetic interaction ($\bar{n}c_0=1$ and $\bar{n}c_2=0.1$) as a function of the driving $\alpha/\omega$. 
The Rabi frequency is $\Omega=2$. The background corresponds to values of the tensor magnetization $\langle {F}^2_z \rangle$. 		
The black and white solid lines with dots represent the first-order and second-order phase transitions, respectively.}
\label{Fig3}
\end{figure}
%%%%%%%%%%%%%%%%%%%%%%%%%%%%%%%%%%%%%%%%%%%%%%%%%%%%%%%%

To reveal the extension of the stripe regions more clearly, we study the phase diagram as a function of the driving $\alpha/\omega$. 
The results are shown in Fig.~\ref{Fig3} 
for $\Omega=2$. 
It is clear from Fig.~\ref{Fig3} that without the driving ($\alpha/\omega =0$) the S2 phase exists in an extremely narrow region of  $\varepsilon$. 
This leads to a challenge for its experimental implementation.  
The upper boundary of the S2 phase corresponds to the degeneracy of three minima of the single-particle dispersion, i.e., $E_{-k_{m}}= E_{k=0}=E_{k_{m}}$, and beyond the boundary $E_{k=0}$ becomes the lowest one such that ground state is the ZM phase. 
Below the boundary, we have $E_{-k_{m}}=E_{k_{m}}< E_{k=0}$. 
For low $\Omega$,  the spinors at $\pm k_{m}$ have $\theta_1=0$ and $\varphi=\pi/2$ and the spinor at $k=0$ has $\theta_2=0$.  
The general wave function becomes
\begin{equation}
\Psi=\sqrt{\bar{n}} \begin{pmatrix}
C_-e^{-ikx} \\ -C_0 \\ C_+e^{ikx}
\end{pmatrix},
\label{Approx}
\end{equation}
which is a good approximation for low $\Omega$. By using Eq.~(\ref{Approx}), we find that 
the antiferromagnetic energy  can be minimized as $\langle E \rangle_{c_2}=0$  in both the S1 phase ($|C_{\pm}|=1/\sqrt{2}, C_0=0$) and the S2 phase 
($|C_-|=|C_+|<1/\sqrt{2}$, $C_0\neq 0$, $\Theta=\pi$)  . 
However, the S2 phase is not a minimization of the quadratic Zeeman energy  $\langle E \rangle_{\varepsilon}=\varepsilon\bar{n} ( |C_-|^2+|C_+|^2 )$  for 
$\varepsilon<0$,  so the ground state is the S1 phase. A dominant Rabi frequency $\Omega$ causes a small deviation $\theta_1$ from zero,  i.e., $\theta_1=\delta\theta$, where $\delta\theta>0$ is a very small quantity.  
This term leads to  $\langle E \rangle_{c_2}=8c_2\bar{n}^2|C_+|^4(\delta \theta)^2$ for both the S1 and S2 phases.  
Considering the S1 phase with $|C_+|^2=1/2$ and the S2 phase with $|C_+|^2<1/2$, this extra antiferromagnetic energy prefers the S2 phase as the ground state if the quadratic Zeeman energy is weak. 
If the quadratic Zeeman energy exceeds this extra energy, the S1 phase is back as the ground state. 
Since the extra energy is a small quantity of second order, the S2 ground state exists in a very small $\varepsilon$ domain.

In the presence of the driving, the region of the S2 phase is dramatically extended  around $\varepsilon=0$ (see Fig.~\ref{Fig3}). 
The upward shift of the region is due to the Bessel-function-modulated Rabi frequency.  
As the driving $\alpha/\omega$ increasing from zeros, $\Omega J_0(\alpha/\omega)$ decreases towards zero. 
As shown in Fig.~\ref{Fig2}, for a small $\Omega$, the S2 phase is located around $\varepsilon=0$.  
The dramatic expansion of the existence area is the  consequence of the Floquet-induced spin-exchange interaction.  
The S2 phase can greatly minimize the spin-exchange-interaction energy, which can be easily seen from the approximate wave function for low $\Omega$ in Eq.~(\ref{Approx}). 
With the wave function, the spin-exchange-interaction energy
 becomes $\langle E \rangle_{c_f}=2c_f\bar{n}^2 |C_-||C_+||C_0|^2\cos(\Theta)$. The S2 phase, having $0<|C_-|=|C_+|<1/\sqrt{2}$ and $\Theta=\pi$,  
minimizes the spin-exchange energy. 
Other phases, such as the ZM phase ($C_0=1$), the PW phase ($C_0=0$, $|C_+|+|C_-|=1$), and the S1 phase ($C_0=0, |C_{\pm}|=1/\sqrt{2}$),  
lead to $\langle E \rangle_{c_f}=0$, 
so that the Floquet-induced spin-exchange energy cannot be minimized. 
Meanwhile, the S2 phase also minimizes the antiferromagnetic interaction energy, $\langle E \rangle_{c_2}=c_2\bar{n}^2/2  (|C_-|^2-|C_+|^2)^2 +c_2\bar{n}^2[|C_-|^2|C_0|^2+|C_+|^2|C_0|^2+2|C_-||C_+||C_0|^2\cos(\Theta)]=0$. 
The only obstacle to the existence of the S2 phase is the quadratic Zeeman energy $\langle E \rangle_{\varepsilon}=\varepsilon \bar{n}( |C_-|^2+|C_+|^2  )$.  If $\varepsilon>0$, the quadratic Zeeman energy prefers the ZM phase, and when $\varepsilon<0$, it prefers the S1 phase. 
Therefore, the competition between the Floquet-induced spin-exchange interaction and the quadratic Zeeman field leads to the existence region for the S2 phase, which is dramatically extended in comparison with the usual case with $\alpha/\omega=0$. 
The S2-ZM (white lines with dots) and S2-S1 (black lines with dots) transition lines oscillate as a function of  $\alpha/\omega$. 
It is noted that the maxima of the transition lines correspond to the zeros of $J_0( \alpha/\omega)$; therefore, the oscillations come from $\Omega J_0(\alpha/\omega)$.  
It is also interesting that without the driving the S2 phase always exists in the negative-$\varepsilon$ area, 
but with the driving, it can exist even in positive-$\varepsilon$ areas.

\subsection{Ferromagnetic interactions}

\begin{figure}[t]
\includegraphics[width=3.2in]{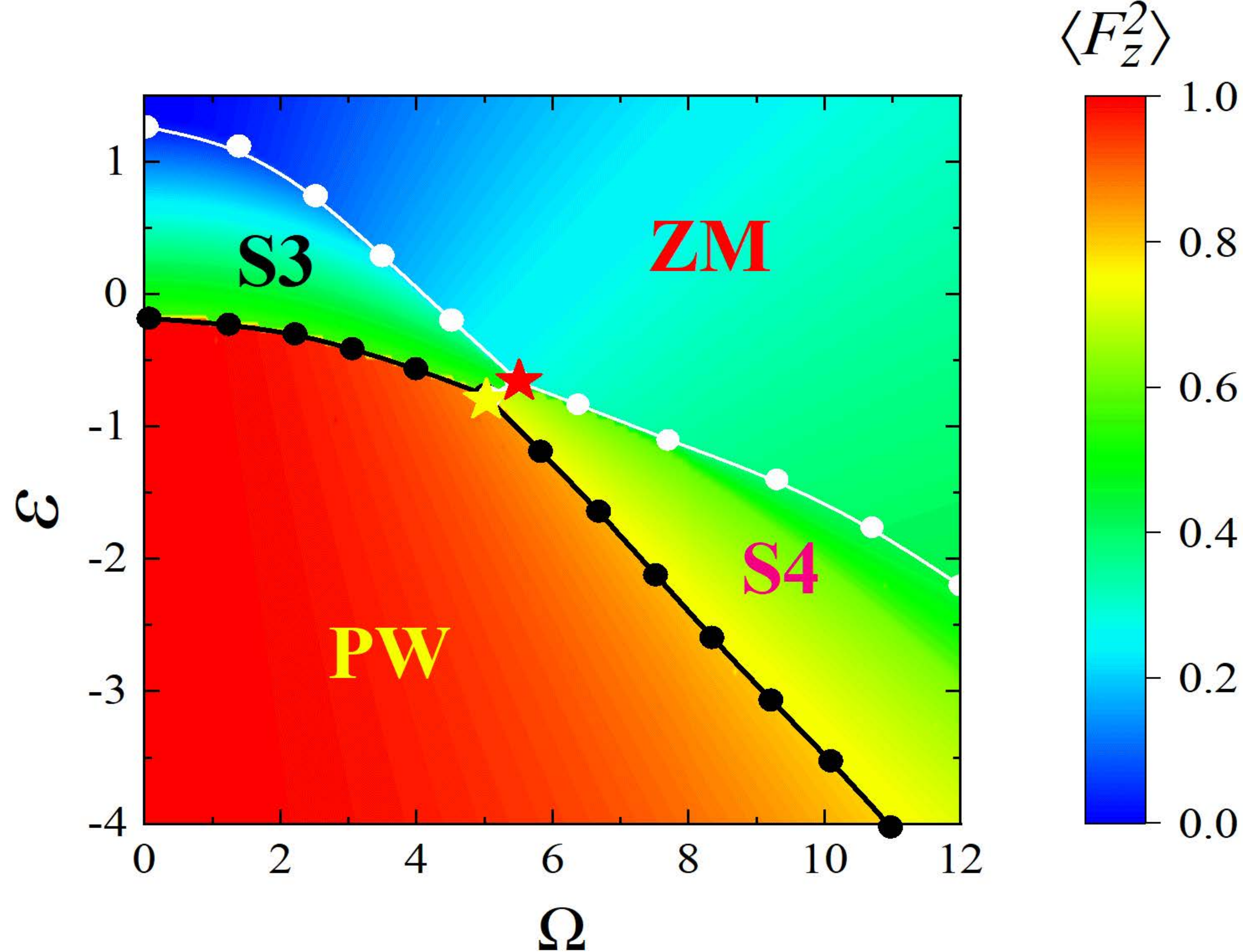}
\caption{Quantum ground-state phase diagram of a spin-orbit-coupled Floquet spinor BEC with a ferromagnetic spin-spin interaction $\bar{n}c_0=1$ and $\bar{n}c_2=-0.5$. The background corresponds to values of the tensor magnetization $\langle {F}^2_z \rangle$. 
The black and white solid lines with dots represent the first-order and second-order phase transitions, respectively. 
The different tricritical points are denoted by red and yellow stars. The driving is $\alpha/\omega=1.6$ [$J_0(\alpha/\omega)=0.455$ and $J_0(2\alpha/\omega)=-0.320$]. }
\label{Fig4}
\end{figure}

The ferromagnetic interaction is $c_2<0$. We consider $c_2/c_0=-0.5$, which is typical of $^7$Li atoms~\cite{Martone2016}.  
Figure~\ref{Fig4} demonstrates the phase diagram for ferromagnetic interactions with driving $\alpha/\omega=1.6$.  
In the low-$\Omega$ region, there is a third stripe phase, which is labeled S3 in Fig.~\ref{Fig4}. 
It has $|C_-|=|C_+|\ne 0,  |C_0| \ne 0$, and $\Theta=0$. 
Using the approximate wave function in Eq.~(\ref{Approx}), we know that the S3 phase minimizes only the second term in the ferromagnetic interaction energy $\langle E \rangle_{c_2}=c_2\bar{n}^2/2  (|C_-|^2-|C_+|^2)^2 +c_2\bar{n}^2[|C_-|^2|C_0|^2+|C_+|^2|C_0|^2+2|C_-||C_+||C_0|^2\cos(\Theta)]$ ($c_2<0$) and it cannot minimize the first term $c_2\bar{n}^2/2  (|C_-|^2-|C_+|^2)^2$ which is minimized by the PW phase. With the effect of the quadratic Zeeman field, the S3, PW and ZM phases are distributed in the way shown in Fig.~\ref{Fig4}. 
These three phases are similar to those in previous studies~\cite{Yu,Martone2016} (i.e., Fig.~1(b) in~\cite{Yu} and Fig.~2 in~\cite{Martone2016}), 
but with the outstanding feature that every phase exists in a broadened region of $\Omega$ due to the Bessel-function modulation.

%%%%%%%%%%%%%%%%%%%%%%%%%%%%%%%%%%%%%%%%%%%%%%%%%%%%%%%%
%%%%%%%%%%%%%%%%%%%%%%%%%%%%%%%%%%%%%%%%%%%%%%%%%%%%%%%%

Different from the case of $\alpha/\omega=0$ in Refs.~\cite{Yu,Martone2016, Campbell}, 
we find  in the Floquet spinor BEC that there a new stripe phase, which is labeled as S4, exists.  
The S4 phase is located inside the region where the single-particle dispersion has two  energy minima at $\pm k_m$, 
and they are equally occupied by the S4 phase with $|C_{\pm}|=1/\sqrt{2}$ and  $C_0=0$. 
This condition is exactly the same as the S1 phase with antiferromagnetic interactions. 
Nevertheless, the S1 phase exists in the low-$\Omega $ region (see Fig.~\ref{Fig2}), while the S4 phase is in the high-$\Omega$ region (see Fig.~\ref{Fig4}). Such a difference in the existence region related to $\Omega$ leads to different features in the S4 phase.  
With $C_0=0$, the minimization of the ferromagnetic energy  and the Floquet-induced energy demonstrated in Eq.~(\ref{Twoenergy}) leads to $|C_-|=0$ or $|C_+|=0$ for low $\Omega$ ($\theta_1\approx 0$). 
In this case, the ground state is the PW phase with $\left\langle F^2_z\right\rangle\approx 1$, as shown in Fig.~\ref{Fig4}. 
For high $\Omega$, one may have $\theta_1\ne 0$ and $(1+c_f/c_2) \sin^2(2\theta_1)-2\cos^4\theta_1>0$.
The minimization of 
$\langle E \rangle_{c_2} +\langle E \rangle_{c_f}$ requires $|C_{\pm}|=1/\sqrt{2}$, so that the ground state is the S4 phase. 
Due to the existence of the S4 phase, there are two tricritical points, labeled by stars in Fig.~\ref{Fig4}.

%%%%%%%%%%%%%%%%%%%%%%%%%%%%%%%%%%%%%%%%%%%%%%%%%%%%%%%%
\begin{figure}[b]
\includegraphics[ width=3.2in]{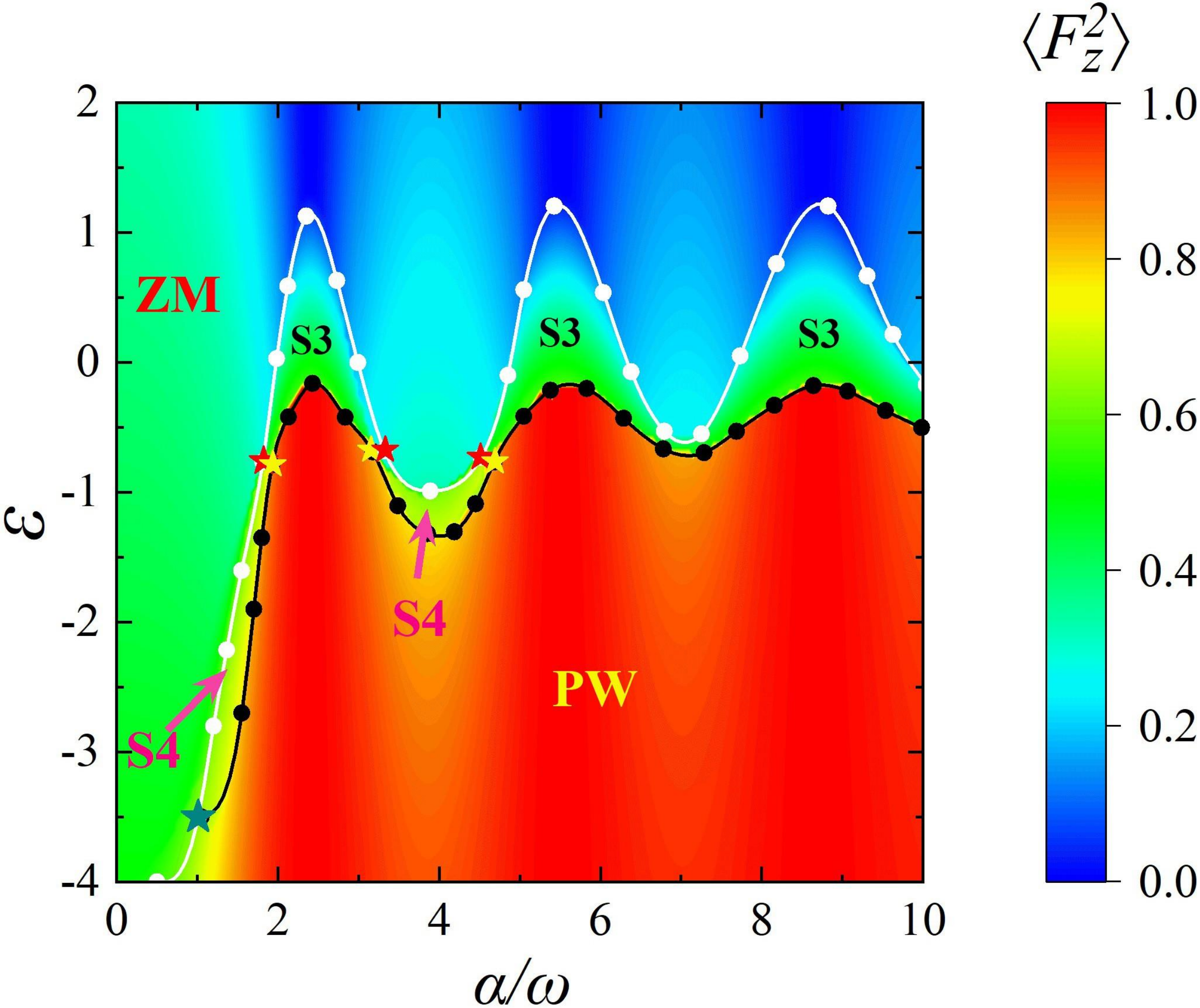}
\caption{Phase diagram in a ferromagnetic interaction ($\bar{n}c_0=1$ and $\bar{n}c_2=-0.5$) as a function of the driving $\alpha/\omega$. The Rabi frequency is $\Omega=8$. The background corresponds to values of the tensor magnetization $\langle {F}^2_z \rangle$. The black and white dotted lines represent the first-order and second-order phase transitions. The different tricritical points are denoted by red and yellow stars.}
\label{Fig5}
\end{figure}
%%%%%%%%%%%%%%%%%%%%%%%%%%%%%%%%%%%%%%%%%%%%%%%%%%%%%%%%

%%%%%%%%%%%%%%%%%%%%%%%%%%%%%%%%%%%%%%%%%%%%%%%%%%%%%%%%
\begin{figure*}[t]
\includegraphics[ width=6in]{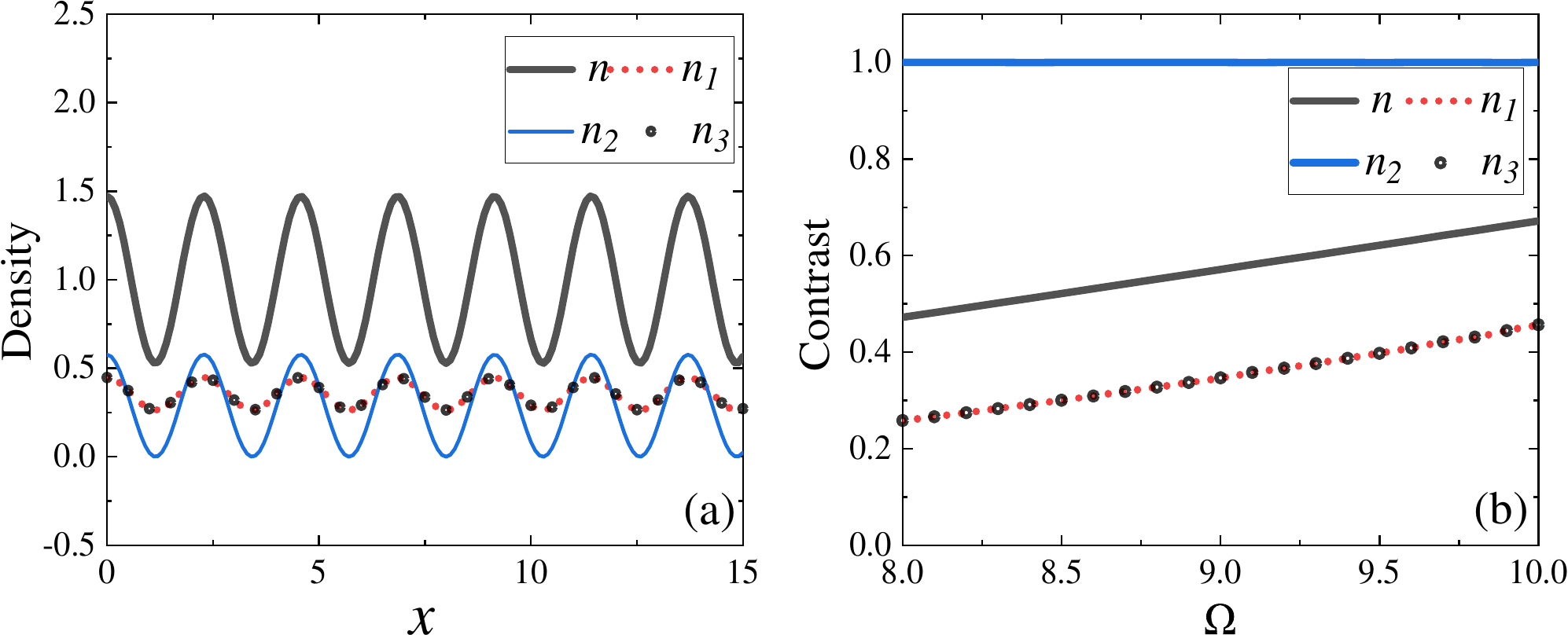}
\caption{Profiles of the S4 phase in a  ferromagnetic interaction ($\bar{n}c_0=1$ and $\bar{n}c_2=-0.5$). 
The driving is $\alpha/\omega=1.6$ ($J_0(\alpha/\omega)=0.455$ and $J_0(2\alpha/\omega)=-0.320$). 
The quadratic Zeeman field is $\varepsilon=-2$. 
(a) Spatial density distributions, $n_i=|\Psi_i|^2$, where $n$ is the total density $n=n_1+n_2+n_3$.  
The Rabi frequency is $\Omega=8$.  
(b) The contrast $(n_{\mathrm{max}}-n_{\mathrm{min}})/(n_{\mathrm{max}}+n_{\mathrm{min}})$ as a function of $\Omega$.}
\label{Fig6}
\end{figure*}
%%%%%%%%%%%%%%%%%%%%%%%%%%%%%%%%%%%%%%%%%%%%%%%%%%%%%%%%

We want to emphasize that without driving ($c_f=0$), the S4 phase cannot exist~\cite{Yu,Martone2016, Campbell}. 
In the absence of driving, Eq.~(\ref{Twoenergy}) becomes $\langle E \rangle_{c_2}   =c_2\bar{n}^2 /2\cos^4\theta_1+
c_2\bar{n}^2|C_{-}|^2|C_{+}|^2\left[ \sin^2(2\theta_1)-2\cos^4\theta_1\right]$. 
For $c_2<0$, the first term $c_2\bar{n}^2 /2\cos^4\theta_1$ prefers $\theta_1=0$. 
According to the second term, the realization of the S4 phase needs a nonzero $\theta_1$ satisfying  $\sin^2(2\theta_1)-2\cos^4\theta_1>0$, 
which can be achieved by increasing the Rabi frequency.
In addition, the negative Rabi coupling energy is also beneficial for lowing the total energy.
However, for the single-particle dispersion with large $\Omega$ the energy minimum at $k=0$ will be
lower than the energy minima at $k=\pm k_{m}$, and the ground state prefers the ZM phase. 
Thus, there is no way for the S4 phase to exist. 
The Floquet-induced interaction has the nature of spin-exchange. 
It has two effects: the spin-exchange interaction causes direct competition with the first term since it prefers $\theta_1=\pi/4$, 
so the three components having equal populations in each spinor; according to Eq.~(\ref{Twoenergy}), the S4 phase requires $ (1+c_f/c_2) \sin^2(2\theta_1)-2\cos^4\theta_1>0$, 
and the positive $c_f/c_2$ as a prefactor also increases the possibility of $\theta_1$ satisfying the requirement.  Therefore, combined effects of  the Rabi coupling and the Floquet-induced interaction makes the existence of the S4 phase possible.

In order to understand how the S4 phase emerges in the presence of driving, 
we analyze the phase diagram as a function of the driving $\alpha/\omega$, which is demonstrated in Fig.~\ref{Fig5}.  
The Rabi frequency is fixed as $\Omega=8$. 
For $\alpha/\omega=0$, the ground state is the ZM phase, as shown in Fig.~\ref{Fig5},  
which is consistent with the results in Refs.~\cite{Yu,Martone2016, Campbell}. 
As $\alpha/\omega$ increasing, the S3, S4 and PW phases appear and have an interesting distribution shown in Fig.~\ref{Fig5}. 
Transition lines (white and black solid lines with dots)  have an oscillating behavior with the maxima matching the zeros of $J_0(\alpha/\omega)$. 
The S3 and S4 phases locate between two transition lines. 
Furthermore, the S4 phase exists in limited regions.  
The change in $\alpha/\omega$ is equivalent to scanning  $\Omega$. 
A high $\alpha/\omega$ leads to $\Omega J_0(\alpha/\omega)$ being confined around zero. 
According to Fig.~\ref{Fig4}, the ground state around $\Omega=0$  is the S3 phase. 
Therefore, for high $\alpha/\omega$ there is no S4 phase anymore (see Fig.~\ref{Fig5}).

In Fig.~\ref{Fig6}(a), we show density distributions $n_i=|\Psi_i|^2$ of a typical S4 state. 
The outstanding feature is that the second component $n_2$ is comparable with the other components $n_1=n_3$. 
This is completely different from the S1 phase with antiferromagnetic interactions, where $n_2 \ll n_1=n_3$. 
This is due to the low-$\Omega$ region for the S1 phase.  
For low $\Omega$, the spinors at $\pm k_{m}$ can be physically approximated as $e^{ik_mx}(\delta^2, \delta, 1)^T$ and $e^{-ik_mx}(1, \delta, \delta^2)^T$, respectively, where $\delta$ is a small quantity. The S1 phase is an equal superposition of the two spinors, and we have $n_1=n_3=1+2\delta^2\cos(2k_{m}x)$ and $n_2=4\delta^2\cos^2(k_{m}x)$. Therefore, the S1 phase has $n_2 \ll n_1=n_3$ and a very low contrast for $n_1$ and $n_2$ which is proportional to a small quantity of second order.  The contrast is defined as $(n_{\mathrm{max}}-n_{\mathrm{min}})/(n_{\mathrm{max}}+n_{\mathrm{min}})$, with $n_{\mathrm{max}}$ ($n_{\mathrm{min}}$) being the density maximum (minimum). 
The low contrast of the S1 phase is unfavorable for experimental observations.  
However, the S4 phase with ferromagnetic interactions exists in the high-$\Omega$ region, and with further help from the Floquet-induced spin-exchange, $\delta$ is not a small quantity anymore. 
Therefore, the contrast of $n_1$ and $n_3$ is obviously high for the S4 phase. 
The advantage of the second component is that its contrast is always maximized (it is equal to 1).  
The dominant occupation in the second component makes it perfect for direct experimental observations. 
In Fig.~\ref{Fig6}(b), we show the contrast in the full $\Omega$ region. 
The contrast of $n_1$ and $n_3$ increases with the increase of $\Omega$, and it is always 1 as expected for the second component $n_2$.

%%%%%%%%%%%%%%%%%%%%%%%%%%%%%%%%%%%%%%%%%%%%%%%%%%%%%%%%
\begin{figure}[t]
\includegraphics[ width=3in]{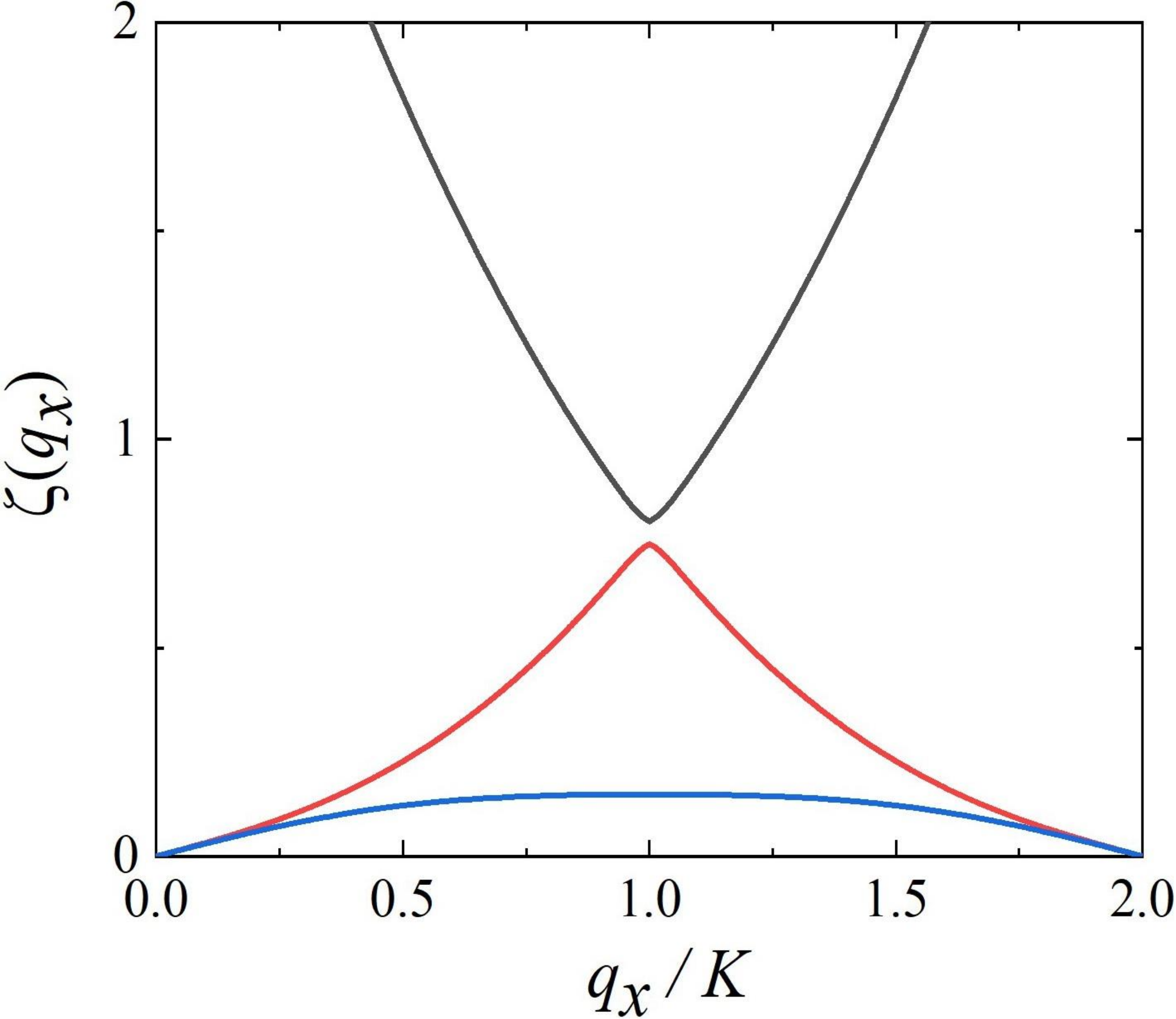}
\caption{Bogoliubov excitation spectrum $\zeta(q_x)$ of a typical S4 state. 
The parameters are $\bar{n}c_0=1$, $\bar{n}c_2=-0.5$, $\Omega=7$, and $\varepsilon=-1.1$.  
The two lowest bands are gapless, corresponding to two Nambu-Goldstone modes.
}
\label{Fig7}
\end{figure}
%%%%%%%%%%%%%%%%%%%%%%%%%%%%%%%%%%%%%%%%%%%%%%%%%%%%%%%%

A closely related topic for ground states is their elementary excitations. 
The excitation spectrum of each phase in typical spin-orbit-coupled spin-1 BECs has been investigated~\cite{Sun,Yu, ChenY}. 
The S4 phase exists only in Floquet spinor BECs, and we study its Bogoliubov excitation.  
The stripe wave function ansatz in Eq.~(\ref{eq:variation}) includes only low-order plane waves. 
It is known that such an ansatz cannot precisely capture Bogoliubov excitation and high-order plane waves should be involved~
\cite{Martone2014,LiY2013,Xiaolong,Lyu,Guanqiangli2021}. 
Therefore, we use the ansatz with high-order modes~\cite{ChenY},
\begin{equation}
\Psi= \sqrt{\bar{n}} \sum_{j=-L}^L e^{ijKx}\begin{pmatrix}
\varphi_1^{(j)} \\  \varphi_2^{(j)} \\   \varphi_3^{(j)}
\end{pmatrix},
\label{stripeexcitation}
\end{equation}
with the normalization condition $\sum_{\sigma,j}|\varphi_\sigma^{(j)}|^2=1$. 
Here, $L$ is the cutoff of the plane waves, and $K$is related to the period of the stripes. 
Spinors $ (\varphi_1^{(j)}, \varphi_2^{(j)},  \varphi_3^{(j)})^T$ and $K$ are determined by minimizing the energy function in Eq.~(\ref{eq:energy}) using Eq.~(\ref{stripeexcitation}). 
In the S4 phase parameter region, we first get the stripe wave function using minimization procedures, 
and then we use the ground state to solve Bogoliubov--de Gennes equation to get the elementary excitation energy $\zeta$~\cite{ChenY}.  
A typical excitation spectrum $\zeta(q_x)$, i.e., the relation between excitation energy $\zeta$ and excitation quasimomentum $q_x$, is demonstrated in Fig.~\ref{Fig7}, in which only the three lowest bands are shown. The size of the Brillouin zone is $2K$, which means that the period of stripes is $\pi/K$. The lowest two bands are gapless, corresponding to two Nambu-Goldstone modes. 
The physical origin of these two gapless modes is the fact that stripes spontaneously break the continuously translational symmetry and gauge symmetry~\cite{LiY2013}.

Finally, we discuss possible experimental observations of our results.  
In the spin-1 spin-orbit-coupled experiment in~\cite{Campbell}, 
the Rabi frequency $\Omega$ and the constant quadratic Zeeman shift $\varepsilon$ are completely tunable. 
$\Omega$ can be experimentally tuned up to $15$ with units of $E_\text{R}$, and $\varepsilon$ can reach $\pm 5E_\text{R}$.  
$\Omega$ and $\varepsilon$ in our study are completely within experimental accessibility. 
In a spin-orbit-coupled degenerate Fermi gas experiment in~\cite{Huang2018}, 
the periodic driving of the frequency of a Raman laser was implemented to reach the high-frequency driving regime.  
We expect that similar to this experiment, our driving can be realized precisely. Under such a high-frequency driving regime, the possible micromotions induced by high-order drivings are safely suppressed, as shown in spin-orbit-coupled experiments~\cite{Jimenez-Garcia,Huang2018}.  
So far, experiment have realized spin-1 spin-orbit coupling in a $^{87}$Rb BEC~\cite{Campbell}. 
The $^{87}$Rb BEC is ferromagnetic but with a very small ratio $c_2/c_0$. Such a small ratio makes the stripe phase S3 exists in a negligible parameter regime~\cite{Martone2016}. In the presence of driving, there is no the new S4 phase for the  $^{87}$Rb BEC.

\section{Conclusion}
\label{conclusion}

Spin-orbit-coupled spin-1 BECs have been realized in experiments. 
Based on the experimental platform, we proposed a spin-orbit-coupled Floquet spinor BEC by periodically driving the quadratic Zeeman field with a high-frequency. 
In the Floquet spinor BEC, the Rabi frequency is modulated by a Bessel function and a Floquet-induced spin-exchange interaction emerges. 
We studied quantum ground-state phase diagram of a spin-orbit-coupled Floquet spinor BEC while considering antiferromagnetic and ferromagnetic spin-spin interactions separately.  
A general result is that due to the Bessel-function modulation, every phase in the diagram can exist in a broadened Rabi frequency region. 
For antiferromagnetic interactions, we found that the existence of a stripe phase can be dramatically extended in the $\varepsilon$ domain due to the Floquet-induced spin-exchange interaction. For ferromagnetic interactions, a different stripe phase was revealed, and its features, 
including high contrast and Bogoliubov excitations, were identified. 
In all previous studies of spin-$1/2$ and spin-1 spin-orbit-coupled BECs, stripes have a very low contrast since they exist in in the low-$\Omega$ regime and the contrast is proportional to the Rabi frequency $\Omega$~\cite{Martone2014}. 
This stripe phase in the Floquet spinor BEC exists in the high-$\Omega$ region and its high contrast favors experimental observations.

\section*{Acknowledgments}
	
We thank Prof. P.~Engels for stimulating discussions. This work is supported by the National Natural Science Foundation of China under Grants No. 11974235 and No. 11774219. 
H.L acknowledges support from Okinawa  Institute of Science and Technology Graduate University.

\bibliography{SOCSpin1Foquet}

\begin{thebibliography}{54}%
\makeatletter
\providecommand \@ifxundefined [1]{%
 \@ifx{#1\undefined}
}%
\providecommand \@ifnum [1]{%
 \ifnum #1\expandafter \@firstoftwo
 \else \expandafter \@secondoftwo
 \fi
}%
\providecommand \@ifx [1]{%
 \ifx #1\expandafter \@firstoftwo
 \else \expandafter \@secondoftwo
 \fi
}%
\providecommand \natexlab [1]{#1}%
\providecommand \enquote  [1]{``#1''}%
\providecommand \bibnamefont  [1]{#1}%
\providecommand \bibfnamefont [1]{#1}%
\providecommand \citenamefont [1]{#1}%
\providecommand \href@noop [0]{\@secondoftwo}%
\providecommand \href [0]{\begingroup \@sanitize@url \@href}%
\providecommand \@href[1]{\@@startlink{#1}\@@href}%
\providecommand \@@href[1]{\endgroup#1\@@endlink}%
\providecommand \@sanitize@url [0]{\catcode `\\12\catcode `\$12\catcode
  `\&12\catcode `\#12\catcode `\^12\catcode `\_12\catcode `\%12\relax}%
\providecommand \@@startlink[1]{}%
\providecommand \@@endlink[0]{}%
\providecommand \url  [0]{\begingroup\@sanitize@url \@url }%
\providecommand \@url [1]{\endgroup\@href {#1}{\urlprefix }}%
\providecommand \urlprefix  [0]{URL }%
\providecommand \Eprint [0]{\href }%
\providecommand \doibase [0]{https://doi.org/}%
\providecommand \selectlanguage [0]{\@gobble}%
\providecommand \bibinfo  [0]{\@secondoftwo}%
\providecommand \bibfield  [0]{\@secondoftwo}%
\providecommand \translation [1]{[#1]}%
\providecommand \BibitemOpen [0]{}%
\providecommand \bibitemStop [0]{}%
\providecommand \bibitemNoStop [0]{.\EOS\space}%
\providecommand \EOS [0]{\spacefactor3000\relax}%
\providecommand \BibitemShut  [1]{\csname bibitem#1\endcsname}%
\let\auto@bib@innerbib\@empty
%</preamble>
\bibitem [{\citenamefont {Goldman}\ \emph {et~al.}(2014)\citenamefont
  {Goldman}, \citenamefont {Juzeli{\={u}}nas}, \citenamefont {\"Ohberg},\ and\
  \citenamefont {Spielman}}]{Goldman}%
  \BibitemOpen
  \bibfield  {author} {\bibinfo {author} {\bibfnamefont {N.}~\bibnamefont
  {Goldman}}, \bibinfo {author} {\bibfnamefont {G.}~\bibnamefont
  {Juzeli{\={u}}nas}}, \bibinfo {author} {\bibfnamefont {P.}~\bibnamefont
  {\"Ohberg}},\ and\ \bibinfo {author} {\bibfnamefont {I.~B.}\ \bibnamefont
  {Spielman}},\ }\bibfield  {title} {\bibinfo {title} {Light-induced gauge
  fields for ultracold atoms},\ }\href
  {https://doi.org/10.1088/0034-4885/77/12/126401} {\bibfield  {journal}
  {\bibinfo  {journal} {Rep. Prog. Phys.}\ }\textbf {\bibinfo {volume} {77}},\
  \bibinfo {pages} {126401} (\bibinfo {year} {2014})}\BibitemShut {NoStop}%
\bibitem [{\citenamefont {Galitski}\ and\ \citenamefont
  {Spielman}(2013)}]{Galitski}%
  \BibitemOpen
  \bibfield  {author} {\bibinfo {author} {\bibfnamefont {V.}~\bibnamefont
  {Galitski}}\ and\ \bibinfo {author} {\bibfnamefont {I.~B.}\ \bibnamefont
  {Spielman}},\ }\bibfield  {title} {\bibinfo {title} {Spin–orbit coupling in
  quantum gases},\ }\href {https://doi.org/10.1038/nature11841} {\bibfield
  {journal} {\bibinfo  {journal} {Nature}\ }\textbf {\bibinfo {volume} {494}},\
  \bibinfo {pages} {49} (\bibinfo {year} {2013})}\BibitemShut {NoStop}%
\bibitem [{\citenamefont {Zhai}(2015)}]{Zhai2015}%
  \BibitemOpen
  \bibfield  {author} {\bibinfo {author} {\bibfnamefont {H.}~\bibnamefont
  {Zhai}},\ }\bibfield  {title} {\bibinfo {title} {Degenerate quantum gases
  with spin{\textendash}orbit coupling: a review},\ }\href
  {https://doi.org/10.1088/0034-4885/78/2/026001} {\bibfield  {journal}
  {\bibinfo  {journal} {Rep. Prog. Phys.}\ }\textbf {\bibinfo {volume} {78}},\
  \bibinfo {pages} {026001} (\bibinfo {year} {2015})}\BibitemShut {NoStop}%
\bibitem [{\citenamefont {Zhang}\ \emph {et~al.}(2016)\citenamefont {Zhang},
  \citenamefont {Mossman}, \citenamefont {Busch}, \citenamefont {Engels},\ and\
  \citenamefont {Zhang}}]{Zhang2016}%
  \BibitemOpen
  \bibfield  {author} {\bibinfo {author} {\bibfnamefont {Y.}~\bibnamefont
  {Zhang}}, \bibinfo {author} {\bibfnamefont {M.~E.}\ \bibnamefont {Mossman}},
  \bibinfo {author} {\bibfnamefont {T.}~\bibnamefont {Busch}}, \bibinfo
  {author} {\bibfnamefont {P.}~\bibnamefont {Engels}},\ and\ \bibinfo {author}
  {\bibfnamefont {C.}~\bibnamefont {Zhang}},\ }\bibfield  {title} {\bibinfo
  {title} {Properties of spin–orbit-coupled {B}ose–{E}instein
  condensates},\ }\href {https://doi.org/10.1007/s11467-016-0560-y} {\bibfield
  {journal} {\bibinfo  {journal} {Front. Phys.}\ }\textbf {\bibinfo {volume}
  {11}},\ \bibinfo {pages} {118103} (\bibinfo {year} {2016})}\BibitemShut
  {NoStop}%
\bibitem [{\citenamefont {Lin}\ \emph {et~al.}(2011)\citenamefont {Lin},
  \citenamefont {Jim\'enez-García},\ and\ \citenamefont {Spielman}}]{Lin}%
  \BibitemOpen
  \bibfield  {author} {\bibinfo {author} {\bibfnamefont {Y.-J.}\ \bibnamefont
  {Lin}}, \bibinfo {author} {\bibfnamefont {K.}~\bibnamefont
  {Jim\'enez-García}},\ and\ \bibinfo {author} {\bibfnamefont {I.~B.}\
  \bibnamefont {Spielman}},\ }\bibfield  {title} {\bibinfo {title}
  {Spin--orbit-coupled {B}ose–{E}instein condensates},\ }\href
  {https://doi.org/10.1038/nature09887} {\bibfield  {journal} {\bibinfo
  {journal} {Nature}\ }\textbf {\bibinfo {volume} {471}},\ \bibinfo {pages}
  {83} (\bibinfo {year} {2011})}\BibitemShut {NoStop}%
\bibitem [{\citenamefont {Wang}\ \emph {et~al.}(2012)\citenamefont {Wang},
  \citenamefont {Yu}, \citenamefont {Fu}, \citenamefont {Miao}, \citenamefont
  {Huang}, \citenamefont {Chai}, \citenamefont {Zhai},\ and\ \citenamefont
  {Zhang}}]{WangP}%
  \BibitemOpen
  \bibfield  {author} {\bibinfo {author} {\bibfnamefont {P.}~\bibnamefont
  {Wang}}, \bibinfo {author} {\bibfnamefont {Z.-Q.}\ \bibnamefont {Yu}},
  \bibinfo {author} {\bibfnamefont {Z.}~\bibnamefont {Fu}}, \bibinfo {author}
  {\bibfnamefont {J.}~\bibnamefont {Miao}}, \bibinfo {author} {\bibfnamefont
  {L.}~\bibnamefont {Huang}}, \bibinfo {author} {\bibfnamefont
  {S.}~\bibnamefont {Chai}}, \bibinfo {author} {\bibfnamefont {H.}~\bibnamefont
  {Zhai}},\ and\ \bibinfo {author} {\bibfnamefont {J.}~\bibnamefont {Zhang}},\
  }\bibfield  {title} {\bibinfo {title} {Spin-{O}rbit {C}oupled {D}egenerate
  {F}ermi {G}ases},\ }\href {https://doi.org/10.1103/PhysRevLett.109.095301}
  {\bibfield  {journal} {\bibinfo  {journal} {Phys. Rev. Lett.}\ }\textbf
  {\bibinfo {volume} {109}},\ \bibinfo {pages} {095301} (\bibinfo {year}
  {2012})}\BibitemShut {NoStop}%
\bibitem [{\citenamefont {Cheuk}\ \emph {et~al.}(2012)\citenamefont {Cheuk},
  \citenamefont {Sommer}, \citenamefont {Hadzibabic}, \citenamefont {Yefsah},
  \citenamefont {Bakr},\ and\ \citenamefont {Zwierlein}}]{Cheuk}%
  \BibitemOpen
  \bibfield  {author} {\bibinfo {author} {\bibfnamefont {L.~W.}\ \bibnamefont
  {Cheuk}}, \bibinfo {author} {\bibfnamefont {A.~T.}\ \bibnamefont {Sommer}},
  \bibinfo {author} {\bibfnamefont {Z.}~\bibnamefont {Hadzibabic}}, \bibinfo
  {author} {\bibfnamefont {T.}~\bibnamefont {Yefsah}}, \bibinfo {author}
  {\bibfnamefont {W.~S.}\ \bibnamefont {Bakr}},\ and\ \bibinfo {author}
  {\bibfnamefont {M.~W.}\ \bibnamefont {Zwierlein}},\ }\bibfield  {title}
  {\bibinfo {title} {Spin-{I}njection {S}pectroscopy of a {S}pin-{O}rbit
  {C}oupled {F}ermi {G}as},\ }\href
  {https://doi.org/10.1103/PhysRevLett.109.095302} {\bibfield  {journal}
  {\bibinfo  {journal} {Phys. Rev. Lett.}\ }\textbf {\bibinfo {volume} {109}},\
  \bibinfo {pages} {095302} (\bibinfo {year} {2012})}\BibitemShut {NoStop}%
\bibitem [{\citenamefont {Ji}\ \emph {et~al.}(2015)\citenamefont {Ji},
  \citenamefont {Zhang}, \citenamefont {Xu}, \citenamefont {Wu}, \citenamefont
  {Deng}, \citenamefont {Chen},\ and\ \citenamefont {Pan}}]{JiS}%
  \BibitemOpen
  \bibfield  {author} {\bibinfo {author} {\bibfnamefont {S.-C.}\ \bibnamefont
  {Ji}}, \bibinfo {author} {\bibfnamefont {L.}~\bibnamefont {Zhang}}, \bibinfo
  {author} {\bibfnamefont {X.-T.}\ \bibnamefont {Xu}}, \bibinfo {author}
  {\bibfnamefont {Z.}~\bibnamefont {Wu}}, \bibinfo {author} {\bibfnamefont
  {Y.}~\bibnamefont {Deng}}, \bibinfo {author} {\bibfnamefont {S.}~\bibnamefont
  {Chen}},\ and\ \bibinfo {author} {\bibfnamefont {J.-W.}\ \bibnamefont
  {Pan}},\ }\bibfield  {title} {\bibinfo {title} {Softening of {R}oton and
  {P}honon {M}odes in a {B}ose-{E}instein {C}ondensate with {S}pin-{O}rbit
  {C}oupling},\ }\href {https://doi.org/10.1103/PhysRevLett.114.105301}
  {\bibfield  {journal} {\bibinfo  {journal} {Phys. Rev. Lett.}\ }\textbf
  {\bibinfo {volume} {114}},\ \bibinfo {pages} {105301} (\bibinfo {year}
  {2015})}\BibitemShut {NoStop}%
\bibitem [{\citenamefont {Wu}\ \emph {et~al.}(2016)\citenamefont {Wu},
  \citenamefont {Zhang}, \citenamefont {Sun}, \citenamefont {Xu}, \citenamefont
  {Wang}, \citenamefont {Ji}, \citenamefont {Deng}, \citenamefont {Chen},
  \citenamefont {Liu},\ and\ \citenamefont {Pan}}]{WuZ}%
  \BibitemOpen
  \bibfield  {author} {\bibinfo {author} {\bibfnamefont {Z.}~\bibnamefont
  {Wu}}, \bibinfo {author} {\bibfnamefont {L.}~\bibnamefont {Zhang}}, \bibinfo
  {author} {\bibfnamefont {W.}~\bibnamefont {Sun}}, \bibinfo {author}
  {\bibfnamefont {X.-T.}\ \bibnamefont {Xu}}, \bibinfo {author} {\bibfnamefont
  {B.-Z.}\ \bibnamefont {Wang}}, \bibinfo {author} {\bibfnamefont {S.-C.}\
  \bibnamefont {Ji}}, \bibinfo {author} {\bibfnamefont {Y.}~\bibnamefont
  {Deng}}, \bibinfo {author} {\bibfnamefont {S.}~\bibnamefont {Chen}}, \bibinfo
  {author} {\bibfnamefont {X.-J.}\ \bibnamefont {Liu}},\ and\ \bibinfo {author}
  {\bibfnamefont {J.-W.}\ \bibnamefont {Pan}},\ }\bibfield  {title} {\bibinfo
  {title} {Realization of two-dimensional spin-orbit coupling for
  {Bose}-{Einstein} condensates},\ }\href
  {https://doi.org/10.1126/science.aaf6689} {\bibfield  {journal} {\bibinfo
  {journal} {Science}\ }\textbf {\bibinfo {volume} {354}},\ \bibinfo {pages}
  {83} (\bibinfo {year} {2016})}\BibitemShut {NoStop}%
\bibitem [{\citenamefont {Huang}\ \emph {et~al.}(2016)\citenamefont {Huang},
  \citenamefont {Meng}, \citenamefont {Wang}, \citenamefont {Peng},
  \citenamefont {Zhang}, \citenamefont {Chen}, \citenamefont {Li},
  \citenamefont {Zhou},\ and\ \citenamefont {Zhang}}]{HuangL}%
  \BibitemOpen
  \bibfield  {author} {\bibinfo {author} {\bibfnamefont {L.}~\bibnamefont
  {Huang}}, \bibinfo {author} {\bibfnamefont {Z.}~\bibnamefont {Meng}},
  \bibinfo {author} {\bibfnamefont {P.}~\bibnamefont {Wang}}, \bibinfo {author}
  {\bibfnamefont {P.}~\bibnamefont {Peng}}, \bibinfo {author} {\bibfnamefont
  {S.-L.}\ \bibnamefont {Zhang}}, \bibinfo {author} {\bibfnamefont
  {L.}~\bibnamefont {Chen}}, \bibinfo {author} {\bibfnamefont {D.}~\bibnamefont
  {Li}}, \bibinfo {author} {\bibfnamefont {Q.}~\bibnamefont {Zhou}},\ and\
  \bibinfo {author} {\bibfnamefont {J.}~\bibnamefont {Zhang}},\ }\bibfield
  {title} {\bibinfo {title} {Experimental realization of two-dimensional
  synthetic spin–orbit coupling in ultracold {F}ermi gases},\ }\href
  {https://doi.org/10.1038/nphys3672} {\bibfield  {journal} {\bibinfo
  {journal} {Nat. Phys.}\ }\textbf {\bibinfo {volume} {12}},\ \bibinfo {pages}
  {540} (\bibinfo {year} {2016})}\BibitemShut {NoStop}%
\bibitem [{\citenamefont {Mossman}\ \emph {et~al.}(2019)\citenamefont
  {Mossman}, \citenamefont {Hou}, \citenamefont {Luo}, \citenamefont {Zhang},\
  and\ \citenamefont {Engels}}]{Mossman}%
  \BibitemOpen
  \bibfield  {author} {\bibinfo {author} {\bibfnamefont {M.~E.}\ \bibnamefont
  {Mossman}}, \bibinfo {author} {\bibfnamefont {J.}~\bibnamefont {Hou}},
  \bibinfo {author} {\bibfnamefont {X.-W.}\ \bibnamefont {Luo}}, \bibinfo
  {author} {\bibfnamefont {C.}~\bibnamefont {Zhang}},\ and\ \bibinfo {author}
  {\bibfnamefont {P.}~\bibnamefont {Engels}},\ }\bibfield  {title} {\bibinfo
  {title} {Experimental realization of a non-magnetic one-way spin switch},\
  }\href {https://doi.org/10.1038/s41467-019-11210-z} {\bibfield  {journal}
  {\bibinfo  {journal} {Nat. Commun.}\ }\textbf {\bibinfo {volume} {10}},\
  \bibinfo {pages} {3381} (\bibinfo {year} {2019})}\BibitemShut {NoStop}%
\bibitem [{\citenamefont {Vald\'es-Curiel}\ \emph {et~al.}(2021)\citenamefont
  {Vald\'es-Curiel}, \citenamefont {Trypogeorgos}, \citenamefont {Liang},
  \citenamefont {Anderson},\ and\ \citenamefont {Spielman}}]{Valdes}%
  \BibitemOpen
  \bibfield  {author} {\bibinfo {author} {\bibfnamefont {A.}~\bibnamefont
  {Vald\'es-Curiel}}, \bibinfo {author} {\bibfnamefont {D.}~\bibnamefont
  {Trypogeorgos}}, \bibinfo {author} {\bibfnamefont {Q.-Y.}\ \bibnamefont
  {Liang}}, \bibinfo {author} {\bibfnamefont {R.~P.}\ \bibnamefont
  {Anderson}},\ and\ \bibinfo {author} {\bibfnamefont {I.~B.}\ \bibnamefont
  {Spielman}},\ }\bibfield  {title} {\bibinfo {title} {Topological features
  without a lattice in {R}ashba spin-orbit coupled atoms},\ }\href
  {https://doi.org/10.1038/s41467-020-20762-4} {\bibfield  {journal} {\bibinfo
  {journal} {Nat. Commun.}\ }\textbf {\bibinfo {volume} {12}},\ \bibinfo
  {pages} {593} (\bibinfo {year} {2021})}\BibitemShut {NoStop}%
\bibitem [{\citenamefont {Fr\"olian}\ \emph {et~al.}(2022)\citenamefont
  {Fr\"olian}, \citenamefont {Chisholm}, \citenamefont {Neri}, \citenamefont
  {Cabrera}, \citenamefont {Ramos}, \citenamefont {Celi},\ and\ \citenamefont
  {Tarruell}}]{Frolian}%
  \BibitemOpen
  \bibfield  {author} {\bibinfo {author} {\bibfnamefont {A.}~\bibnamefont
  {Fr\"olian}}, \bibinfo {author} {\bibfnamefont {C.~S.}\ \bibnamefont
  {Chisholm}}, \bibinfo {author} {\bibfnamefont {E.}~\bibnamefont {Neri}},
  \bibinfo {author} {\bibfnamefont {C.~R.}\ \bibnamefont {Cabrera}}, \bibinfo
  {author} {\bibfnamefont {R.}~\bibnamefont {Ramos}}, \bibinfo {author}
  {\bibfnamefont {A.}~\bibnamefont {Celi}},\ and\ \bibinfo {author}
  {\bibfnamefont {L.}~\bibnamefont {Tarruell}},\ }\bibfield  {title} {\bibinfo
  {title} {Realizing a 1{D} topological gauge theory in an optically dressed
  {B}{E}{C}},\ }\href {https://doi.org/10.1038/s41586-022-04943-3} {\bibfield
  {journal} {\bibinfo  {journal} {Nature}\ }\textbf {\bibinfo {volume} {608}},\
  \bibinfo {pages} {293} (\bibinfo {year} {2022})}\BibitemShut {NoStop}%
\bibitem [{\citenamefont {Wang}\ \emph {et~al.}(2010)\citenamefont {Wang},
  \citenamefont {Gao}, \citenamefont {Jian},\ and\ \citenamefont
  {Zhai}}]{Wang2010}%
  \BibitemOpen
  \bibfield  {author} {\bibinfo {author} {\bibfnamefont {C.}~\bibnamefont
  {Wang}}, \bibinfo {author} {\bibfnamefont {C.}~\bibnamefont {Gao}}, \bibinfo
  {author} {\bibfnamefont {C.-M.}\ \bibnamefont {Jian}},\ and\ \bibinfo
  {author} {\bibfnamefont {H.}~\bibnamefont {Zhai}},\ }\bibfield  {title}
  {\bibinfo {title} {Spin-{O}rbit {C}oupled {S}pinor {B}ose-{E}instein
  {C}ondensates},\ }\href {https://doi.org/10.1103/PhysRevLett.105.160403}
  {\bibfield  {journal} {\bibinfo  {journal} {Phys. Rev. Lett.}\ }\textbf
  {\bibinfo {volume} {105}},\ \bibinfo {pages} {160403} (\bibinfo {year}
  {2010})}\BibitemShut {NoStop}%
\bibitem [{\citenamefont {Wu}\ \emph {et~al.}(2011)\citenamefont {Wu},
  \citenamefont {Mondragon-Shem},\ and\ \citenamefont {Zhou}}]{Wucong2011}%
  \BibitemOpen
  \bibfield  {author} {\bibinfo {author} {\bibfnamefont {C.}~\bibnamefont
  {Wu}}, \bibinfo {author} {\bibfnamefont {I.}~\bibnamefont {Mondragon-Shem}},\
  and\ \bibinfo {author} {\bibfnamefont {X.-F.}\ \bibnamefont {Zhou}},\
  }\bibfield  {title} {\bibinfo {title} {Unconventional {B}ose-{E}instein
  {C}ondensations from {S}pin-{O}rbit {C}oupling},\ }\href
  {https://doi.org/10.1088/0256-307X/28/9/097102} {\bibfield  {journal}
  {\bibinfo  {journal} {Chin. Phys. Lett.}\ }\textbf {\bibinfo {volume} {28}},\
  \bibinfo {pages} {097102} (\bibinfo {year} {2011})}\BibitemShut {NoStop}%
\bibitem [{\citenamefont {Ho}\ and\ \citenamefont {Zhang}(2011)}]{Hotian2011}%
  \BibitemOpen
  \bibfield  {author} {\bibinfo {author} {\bibfnamefont {T.-L.}\ \bibnamefont
  {Ho}}\ and\ \bibinfo {author} {\bibfnamefont {S.}~\bibnamefont {Zhang}},\
  }\bibfield  {title} {\bibinfo {title} {Bose-{E}instein {C}ondensates with
  {S}pin-{O}rbit {I}nteraction},\ }\href
  {https://doi.org/10.1103/PhysRevLett.107.150403} {\bibfield  {journal}
  {\bibinfo  {journal} {Phys. Rev. Lett.}\ }\textbf {\bibinfo {volume} {107}},\
  \bibinfo {pages} {150403} (\bibinfo {year} {2011})}\BibitemShut {NoStop}%
\bibitem [{\citenamefont {Hu}\ \emph {et~al.}(2012)\citenamefont {Hu},
  \citenamefont {Ramachandhran}, \citenamefont {Pu},\ and\ \citenamefont
  {Liu}}]{Hu2012}%
  \BibitemOpen
  \bibfield  {author} {\bibinfo {author} {\bibfnamefont {H.}~\bibnamefont
  {Hu}}, \bibinfo {author} {\bibfnamefont {B.}~\bibnamefont {Ramachandhran}},
  \bibinfo {author} {\bibfnamefont {H.}~\bibnamefont {Pu}},\ and\ \bibinfo
  {author} {\bibfnamefont {X.-J.}\ \bibnamefont {Liu}},\ }\bibfield  {title}
  {\bibinfo {title} {Spin-{O}rbit {C}oupled {W}eakly {I}nteracting
  {B}ose-{E}instein {C}ondensates in {H}armonic {T}raps},\ }\href
  {https://doi.org/10.1103/PhysRevLett.108.010402} {\bibfield  {journal}
  {\bibinfo  {journal} {Phys. Rev. Lett.}\ }\textbf {\bibinfo {volume} {108}},\
  \bibinfo {pages} {010402} (\bibinfo {year} {2012})}\BibitemShut {NoStop}%
\bibitem [{\citenamefont {Zhang}\ \emph {et~al.}(2012)\citenamefont {Zhang},
  \citenamefont {Mao},\ and\ \citenamefont {Zhang}}]{Yongping2012}%
  \BibitemOpen
  \bibfield  {author} {\bibinfo {author} {\bibfnamefont {Y.}~\bibnamefont
  {Zhang}}, \bibinfo {author} {\bibfnamefont {L.}~\bibnamefont {Mao}},\ and\
  \bibinfo {author} {\bibfnamefont {C.}~\bibnamefont {Zhang}},\ }\bibfield
  {title} {\bibinfo {title} {Mean-{F}ield {D}ynamics of {S}pin-{O}rbit
  {C}oupled {B}ose-{E}instein {C}ondensates},\ }\href
  {https://doi.org/10.1103/PhysRevLett.108.035302} {\bibfield  {journal}
  {\bibinfo  {journal} {Phys. Rev. Lett.}\ }\textbf {\bibinfo {volume} {108}},\
  \bibinfo {pages} {035302} (\bibinfo {year} {2012})}\BibitemShut {NoStop}%
\bibitem [{\citenamefont {Li}\ \emph {et~al.}(2012)\citenamefont {Li},
  \citenamefont {Pitaevskii},\ and\ \citenamefont {Stringari}}]{LiY2012}%
  \BibitemOpen
  \bibfield  {author} {\bibinfo {author} {\bibfnamefont {Y.}~\bibnamefont
  {Li}}, \bibinfo {author} {\bibfnamefont {L.~P.}\ \bibnamefont {Pitaevskii}},\
  and\ \bibinfo {author} {\bibfnamefont {S.}~\bibnamefont {Stringari}},\
  }\bibfield  {title} {\bibinfo {title} {Quantum {T}ricriticality and {P}hase
  {T}ransitions in {Sp}in-{O}rbit {C}oupled {B}ose-{E}instein {C}ondensates},\
  }\href {https://doi.org/10.1103/PhysRevLett.108.225301} {\bibfield  {journal}
  {\bibinfo  {journal} {Phys. Rev. Lett.}\ }\textbf {\bibinfo {volume} {108}},\
  \bibinfo {pages} {225301} (\bibinfo {year} {2012})}\BibitemShut {NoStop}%
\bibitem [{\citenamefont {Khamehchi}\ \emph {et~al.}(2014)\citenamefont
  {Khamehchi}, \citenamefont {Zhang}, \citenamefont {Hamner}, \citenamefont
  {Busch},\ and\ \citenamefont {Engels}}]{Khamehchi}%
  \BibitemOpen
  \bibfield  {author} {\bibinfo {author} {\bibfnamefont {M.~A.}\ \bibnamefont
  {Khamehchi}}, \bibinfo {author} {\bibfnamefont {Y.}~\bibnamefont {Zhang}},
  \bibinfo {author} {\bibfnamefont {C.}~\bibnamefont {Hamner}}, \bibinfo
  {author} {\bibfnamefont {T.}~\bibnamefont {Busch}},\ and\ \bibinfo {author}
  {\bibfnamefont {P.}~\bibnamefont {Engels}},\ }\bibfield  {title} {\bibinfo
  {title} {Measurement of collective excitations in a spin-orbit-coupled
  {B}ose-{E}instein condensate},\ }\href
  {https://doi.org/10.1103/PhysRevA.90.063624} {\bibfield  {journal} {\bibinfo
  {journal} {Phys. Rev. A}\ }\textbf {\bibinfo {volume} {90}},\ \bibinfo
  {pages} {063624} (\bibinfo {year} {2014})}\BibitemShut {NoStop}%
\bibitem [{\citenamefont {Li}\ \emph {et~al.}(2013)\citenamefont {Li},
  \citenamefont {Martone}, \citenamefont {Pitaevskii},\ and\ \citenamefont
  {Stringari}}]{LiY2013}%
  \BibitemOpen
  \bibfield  {author} {\bibinfo {author} {\bibfnamefont {Y.}~\bibnamefont
  {Li}}, \bibinfo {author} {\bibfnamefont {G.~I.}\ \bibnamefont {Martone}},
  \bibinfo {author} {\bibfnamefont {L.~P.}\ \bibnamefont {Pitaevskii}},\ and\
  \bibinfo {author} {\bibfnamefont {S.}~\bibnamefont {Stringari}},\ }\bibfield
  {title} {\bibinfo {title} {Superstripes and the {E}xcitation {S}pectrum of a
  {S}pin-{O}rbit-{C}oupled {B}ose-{E}instein {C}ondensate},\ }\href
  {https://doi.org/10.1103/PhysRevLett.110.235302} {\bibfield  {journal}
  {\bibinfo  {journal} {Phys. Rev. Lett.}\ }\textbf {\bibinfo {volume} {110}},\
  \bibinfo {pages} {235302} (\bibinfo {year} {2013})}\BibitemShut {NoStop}%
\bibitem [{\citenamefont {Zheng}\ \emph {et~al.}(2013)\citenamefont {Zheng},
  \citenamefont {Yu}, \citenamefont {Cui},\ and\ \citenamefont
  {Zhai}}]{Zheng2013}%
  \BibitemOpen
  \bibfield  {author} {\bibinfo {author} {\bibfnamefont {W.}~\bibnamefont
  {Zheng}}, \bibinfo {author} {\bibfnamefont {Z.-Q.}\ \bibnamefont {Yu}},
  \bibinfo {author} {\bibfnamefont {X.}~\bibnamefont {Cui}},\ and\ \bibinfo
  {author} {\bibfnamefont {H.}~\bibnamefont {Zhai}},\ }\bibfield  {title}
  {\bibinfo {title} {Properties of bose gases with the raman-induced
  spin–orbit coupling},\ }\href
  {https://doi.org/10.1088/0953-4075/46/13/134007} {\ \textbf {\bibinfo
  {volume} {46}},\ \bibinfo {pages} {134007} (\bibinfo {year}
  {2013})}\BibitemShut {NoStop}%
\bibitem [{\citenamefont {Martone}\ \emph {et~al.}(2014)\citenamefont
  {Martone}, \citenamefont {Li},\ and\ \citenamefont
  {Stringari}}]{Martone2014}%
  \BibitemOpen
  \bibfield  {author} {\bibinfo {author} {\bibfnamefont {G.~I.}\ \bibnamefont
  {Martone}}, \bibinfo {author} {\bibfnamefont {Y.}~\bibnamefont {Li}},\ and\
  \bibinfo {author} {\bibfnamefont {S.}~\bibnamefont {Stringari}},\ }\bibfield
  {title} {\bibinfo {title} {Approach for making visible and stable stripes in
  a spin-orbit-coupled bose-einstein superfluid},\ }\href
  {https://doi.org/10.1103/PhysRevA.90.041604} {\bibfield  {journal} {\bibinfo
  {journal} {Phys. Rev. A}\ }\textbf {\bibinfo {volume} {90}},\ \bibinfo
  {pages} {041604} (\bibinfo {year} {2014})}\BibitemShut {NoStop}%
\bibitem [{\citenamefont {Luo}\ and\ \citenamefont {Zhang}(2019)}]{Luo2019}%
  \BibitemOpen
  \bibfield  {author} {\bibinfo {author} {\bibfnamefont {X.-W.}\ \bibnamefont
  {Luo}}\ and\ \bibinfo {author} {\bibfnamefont {C.}~\bibnamefont {Zhang}},\
  }\bibfield  {title} {\bibinfo {title} {Tunable spin-orbit coupling and
  magnetic superstripe phase in a {B}ose-{E}instein condensate},\ }\href
  {https://doi.org/10.1103/PhysRevA.100.063606} {\bibfield  {journal} {\bibinfo
   {journal} {Phys. Rev. A}\ }\textbf {\bibinfo {volume} {100}},\ \bibinfo
  {pages} {063606} (\bibinfo {year} {2019})}\BibitemShut {NoStop}%
\bibitem [{\citenamefont {Bersano}\ \emph {et~al.}(2019)\citenamefont
  {Bersano}, \citenamefont {Hou}, \citenamefont {Mossman}, \citenamefont
  {Gokhroo}, \citenamefont {Luo}, \citenamefont {Sun}, \citenamefont {Zhang},\
  and\ \citenamefont {Engels}}]{Peter2019}%
  \BibitemOpen
  \bibfield  {author} {\bibinfo {author} {\bibfnamefont {T.~M.}\ \bibnamefont
  {Bersano}}, \bibinfo {author} {\bibfnamefont {J.}~\bibnamefont {Hou}},
  \bibinfo {author} {\bibfnamefont {S.}~\bibnamefont {Mossman}}, \bibinfo
  {author} {\bibfnamefont {V.}~\bibnamefont {Gokhroo}}, \bibinfo {author}
  {\bibfnamefont {X.-W.}\ \bibnamefont {Luo}}, \bibinfo {author} {\bibfnamefont
  {K.}~\bibnamefont {Sun}}, \bibinfo {author} {\bibfnamefont {C.}~\bibnamefont
  {Zhang}},\ and\ \bibinfo {author} {\bibfnamefont {P.}~\bibnamefont
  {Engels}},\ }\bibfield  {title} {\bibinfo {title} {Experimental realization
  of a long-lived striped bose-einstein condensate induced by momentum-space
  hopping},\ }\href {https://doi.org/10.1103/PhysRevA.99.051602} {\bibfield
  {journal} {\bibinfo  {journal} {Phys. Rev. A}\ }\textbf {\bibinfo {volume}
  {99}},\ \bibinfo {pages} {051602} (\bibinfo {year} {2019})}\BibitemShut
  {NoStop}%
\bibitem [{\citenamefont {Li}\ \emph {et~al.}(2017)\citenamefont {Li},
  \citenamefont {Lee}, \citenamefont {Huang}, \citenamefont {Burchesky},
  \citenamefont {Shteynas}, \citenamefont {Top}, \citenamefont {Jamison},\ and\
  \citenamefont {Ketterle}}]{LiJR}%
  \BibitemOpen
  \bibfield  {author} {\bibinfo {author} {\bibfnamefont {J.-R.}\ \bibnamefont
  {Li}}, \bibinfo {author} {\bibfnamefont {J.}~\bibnamefont {Lee}}, \bibinfo
  {author} {\bibfnamefont {W.}~\bibnamefont {Huang}}, \bibinfo {author}
  {\bibfnamefont {S.}~\bibnamefont {Burchesky}}, \bibinfo {author}
  {\bibfnamefont {B.}~\bibnamefont {Shteynas}}, \bibinfo {author}
  {\bibfnamefont {F.~{\c{C}}.}\ \bibnamefont {Top}}, \bibinfo {author}
  {\bibfnamefont {A.~O.}\ \bibnamefont {Jamison}},\ and\ \bibinfo {author}
  {\bibfnamefont {W.}~\bibnamefont {Ketterle}},\ }\bibfield  {title} {\bibinfo
  {title} {A stripe phase with supersolid properties in spin–orbit-coupled
  {B}ose–{E}instein condensates},\ }\href
  {https://doi.org/10.1038/nature21431} {\bibfield  {journal} {\bibinfo
  {journal} {Nature}\ }\textbf {\bibinfo {volume} {543}},\ \bibinfo {pages}
  {91} (\bibinfo {year} {2017})}\BibitemShut {NoStop}%
\bibitem [{\citenamefont {Stamper-Kurn}\ and\ \citenamefont
  {Ueda}(2013)}]{Stamper-Kurn}%
  \BibitemOpen
  \bibfield  {author} {\bibinfo {author} {\bibfnamefont {D.~M.}\ \bibnamefont
  {Stamper-Kurn}}\ and\ \bibinfo {author} {\bibfnamefont {M.}~\bibnamefont
  {Ueda}},\ }\bibfield  {title} {\bibinfo {title} {Spinor {B}ose gases:
  {S}ymmetries, magnetism, and quantum dynamics},\ }\href
  {https://doi.org/10.1103/RevModPhys.85.1191} {\bibfield  {journal} {\bibinfo
  {journal} {Rev. Mod. Phys.}\ }\textbf {\bibinfo {volume} {85}},\ \bibinfo
  {pages} {1191} (\bibinfo {year} {2013})}\BibitemShut {NoStop}%
\bibitem [{\citenamefont {Campbell}\ \emph {et~al.}(2016)\citenamefont
  {Campbell}, \citenamefont {Price}, \citenamefont {Putra}, \citenamefont
  {Vald\'es-Curiel}, \citenamefont {Trypogeorgos},\ and\ \citenamefont
  {Spielman}}]{Campbell}%
  \BibitemOpen
  \bibfield  {author} {\bibinfo {author} {\bibfnamefont {D.~L.}\ \bibnamefont
  {Campbell}}, \bibinfo {author} {\bibfnamefont {R.~M.}\ \bibnamefont {Price}},
  \bibinfo {author} {\bibfnamefont {A.}~\bibnamefont {Putra}}, \bibinfo
  {author} {\bibfnamefont {A.}~\bibnamefont {Vald\'es-Curiel}}, \bibinfo
  {author} {\bibfnamefont {D.}~\bibnamefont {Trypogeorgos}},\ and\ \bibinfo
  {author} {\bibfnamefont {I.~B.}\ \bibnamefont {Spielman}},\ }\bibfield
  {title} {\bibinfo {title} {Magnetic phases of spin-1 spin–orbit-coupled
  {B}ose gases},\ }\href {https://doi.org/10.1038/ncomms10897} {\bibfield
  {journal} {\bibinfo  {journal} {Nat. Commun.}\ }\textbf {\bibinfo {volume}
  {7}},\ \bibinfo {pages} {10897} (\bibinfo {year} {2016})}\BibitemShut
  {NoStop}%
\bibitem [{\citenamefont {Lan}\ and\ \citenamefont {\"Ohberg}(2014)}]{Lan}%
  \BibitemOpen
  \bibfield  {author} {\bibinfo {author} {\bibfnamefont {Z.}~\bibnamefont
  {Lan}}\ and\ \bibinfo {author} {\bibfnamefont {P.}~\bibnamefont {\"Ohberg}},\
  }\bibfield  {title} {\bibinfo {title} {Raman-dressed spin-1
  spin-orbit-coupled quantum gas},\ }\href
  {https://doi.org/10.1103/PhysRevA.89.023630} {\bibfield  {journal} {\bibinfo
  {journal} {Phys. Rev. A}\ }\textbf {\bibinfo {volume} {89}},\ \bibinfo
  {pages} {023630} (\bibinfo {year} {2014})}\BibitemShut {NoStop}%
\bibitem [{\citenamefont {Sun}\ \emph {et~al.}(2016)\citenamefont {Sun},
  \citenamefont {Qu}, \citenamefont {Xu}, \citenamefont {Zhang},\ and\
  \citenamefont {Zhang}}]{Sun}%
  \BibitemOpen
  \bibfield  {author} {\bibinfo {author} {\bibfnamefont {K.}~\bibnamefont
  {Sun}}, \bibinfo {author} {\bibfnamefont {C.}~\bibnamefont {Qu}}, \bibinfo
  {author} {\bibfnamefont {Y.}~\bibnamefont {Xu}}, \bibinfo {author}
  {\bibfnamefont {Y.}~\bibnamefont {Zhang}},\ and\ \bibinfo {author}
  {\bibfnamefont {C.}~\bibnamefont {Zhang}},\ }\bibfield  {title} {\bibinfo
  {title} {Interacting spin-orbit-coupled spin-1 {B}ose-{E}instein
  condensates},\ }\href {https://doi.org/10.1103/PhysRevA.93.023615} {\bibfield
   {journal} {\bibinfo  {journal} {Phys. Rev. A}\ }\textbf {\bibinfo {volume}
  {93}},\ \bibinfo {pages} {023615} (\bibinfo {year} {2016})}\BibitemShut
  {NoStop}%
\bibitem [{\citenamefont {Yu}(2016)}]{Yu}%
  \BibitemOpen
  \bibfield  {author} {\bibinfo {author} {\bibfnamefont {Z.-Q.}\ \bibnamefont
  {Yu}},\ }\bibfield  {title} {\bibinfo {title} {Phase transitions and
  elementary excitations in spin-1 {B}ose gases with {R}aman-induced spin-orbit
  coupling},\ }\href {https://doi.org/10.1103/PhysRevA.93.033648} {\bibfield
  {journal} {\bibinfo  {journal} {Phys. Rev. A}\ }\textbf {\bibinfo {volume}
  {93}},\ \bibinfo {pages} {033648} (\bibinfo {year} {2016})}\BibitemShut
  {NoStop}%
\bibitem [{\citenamefont {Martone}\ \emph {et~al.}(2016)\citenamefont
  {Martone}, \citenamefont {Pepe}, \citenamefont {Facchi}, \citenamefont
  {Pascazio},\ and\ \citenamefont {Stringari}}]{Martone2016}%
  \BibitemOpen
  \bibfield  {author} {\bibinfo {author} {\bibfnamefont {G.~I.}\ \bibnamefont
  {Martone}}, \bibinfo {author} {\bibfnamefont {F.~V.}\ \bibnamefont {Pepe}},
  \bibinfo {author} {\bibfnamefont {P.}~\bibnamefont {Facchi}}, \bibinfo
  {author} {\bibfnamefont {S.}~\bibnamefont {Pascazio}},\ and\ \bibinfo
  {author} {\bibfnamefont {S.}~\bibnamefont {Stringari}},\ }\bibfield  {title}
  {\bibinfo {title} {Tricriticalities and {Q}uantum {P}hases in
  {S}pin-{O}rbit-{C}oupled {S}pin-1 {B}ose {G}ases},\ }\href
  {https://doi.org/10.1103/PhysRevLett.117.125301} {\bibfield  {journal}
  {\bibinfo  {journal} {Phys. Rev. Lett.}\ }\textbf {\bibinfo {volume} {117}},\
  \bibinfo {pages} {125301} (\bibinfo {year} {2016})}\BibitemShut {NoStop}%
\bibitem [{\citenamefont {Chen}\ \emph {et~al.}(2022)\citenamefont {Chen},
  \citenamefont {Lyu}, \citenamefont {Xu},\ and\ \citenamefont
  {Zhang}}]{ChenY}%
  \BibitemOpen
  \bibfield  {author} {\bibinfo {author} {\bibfnamefont {Y.}~\bibnamefont
  {Chen}}, \bibinfo {author} {\bibfnamefont {H.}~\bibnamefont {Lyu}}, \bibinfo
  {author} {\bibfnamefont {Y.}~\bibnamefont {Xu}},\ and\ \bibinfo {author}
  {\bibfnamefont {Y.}~\bibnamefont {Zhang}},\ }\bibfield  {title} {\bibinfo
  {title} {Elementary excitations in a spin–orbit-coupled spin-1
  {B}ose–{E}instein condensate},\ }\href
  {https://doi.org/10.1088/1367-2630/ac7fb1} {\bibfield  {journal} {\bibinfo
  {journal} {New J. Phys.}\ }\textbf {\bibinfo {volume} {24}},\ \bibinfo
  {pages} {073041} (\bibinfo {year} {2022})}\BibitemShut {NoStop}%
\bibitem [{\citenamefont {Bukov}\ \emph {et~al.}(2015)\citenamefont {Bukov},
  \citenamefont {D'Alessio},\ and\ \citenamefont {Polkovnikov}}]{Bukov}%
  \BibitemOpen
  \bibfield  {author} {\bibinfo {author} {\bibfnamefont {M.}~\bibnamefont
  {Bukov}}, \bibinfo {author} {\bibfnamefont {L.}~\bibnamefont {D'Alessio}},\
  and\ \bibinfo {author} {\bibfnamefont {A.}~\bibnamefont {Polkovnikov}},\
  }\bibfield  {title} {\bibinfo {title} {Universal high-frequency behavior of
  periodically driven systems: from dynamical stabilization to {F}loquet
  engineering},\ }\href {https://doi.org/10.1080/00018732.2015.1055918}
  {\bibfield  {journal} {\bibinfo  {journal} {Adv. Phys.}\ }\textbf {\bibinfo
  {volume} {64}},\ \bibinfo {pages} {139} (\bibinfo {year} {2015})}\BibitemShut
  {NoStop}%
\bibitem [{\citenamefont {Eckardt}(2017)}]{Eckardt}%
  \BibitemOpen
  \bibfield  {author} {\bibinfo {author} {\bibfnamefont {A.}~\bibnamefont
  {Eckardt}},\ }\bibfield  {title} {\bibinfo {title} {Colloquium: {A}tomic
  quantum gases in periodically driven optical lattices},\ }\href
  {https://doi.org/10.1103/RevModPhys.89.011004} {\bibfield  {journal}
  {\bibinfo  {journal} {Rev. Mod. Phys.}\ }\textbf {\bibinfo {volume} {89}},\
  \bibinfo {pages} {011004} (\bibinfo {year} {2017})}\BibitemShut {NoStop}%
\bibitem [{\citenamefont {Oka}\ and\ \citenamefont {Kitamura}(2019)}]{Oka}%
  \BibitemOpen
  \bibfield  {author} {\bibinfo {author} {\bibfnamefont {T.}~\bibnamefont
  {Oka}}\ and\ \bibinfo {author} {\bibfnamefont {S.}~\bibnamefont {Kitamura}},\
  }\bibfield  {title} {\bibinfo {title} {Floquet {E}ngineering of {Q}uantum
  {M}aterials},\ }\href
  {https://doi.org/10.1146/annurev-conmatphys-031218-013423} {\bibfield
  {journal} {\bibinfo  {journal} {Annu. Rev. Condens. Matter Phys.}\ }\textbf
  {\bibinfo {volume} {10}},\ \bibinfo {pages} {387} (\bibinfo {year}
  {2019})}\BibitemShut {NoStop}%
\bibitem [{\citenamefont {Jotzu}\ \emph {et~al.}(2014)\citenamefont {Jotzu},
  \citenamefont {Messer}, \citenamefont {Desbuquois}, \citenamefont {Lebrat},
  \citenamefont {Uehlinger}, \citenamefont {Greif},\ and\ \citenamefont
  {Esslinger}}]{Jotzu}%
  \BibitemOpen
  \bibfield  {author} {\bibinfo {author} {\bibfnamefont {G.}~\bibnamefont
  {Jotzu}}, \bibinfo {author} {\bibfnamefont {M.}~\bibnamefont {Messer}},
  \bibinfo {author} {\bibfnamefont {R.}~\bibnamefont {Desbuquois}}, \bibinfo
  {author} {\bibfnamefont {M.}~\bibnamefont {Lebrat}}, \bibinfo {author}
  {\bibfnamefont {T.}~\bibnamefont {Uehlinger}}, \bibinfo {author}
  {\bibfnamefont {D.}~\bibnamefont {Greif}},\ and\ \bibinfo {author}
  {\bibfnamefont {T.}~\bibnamefont {Esslinger}},\ }\bibfield  {title} {\bibinfo
  {title} {Experimental realization of the topological haldane model with
  ultracold fermions},\ }\href {https://doi.org/10.1038/nature13915} {\bibfield
   {journal} {\bibinfo  {journal} {Nature}\ }\textbf {\bibinfo {volume}
  {515}},\ \bibinfo {pages} {237} (\bibinfo {year} {2014})}\BibitemShut
  {NoStop}%
\bibitem [{\citenamefont {Struck}\ \emph {et~al.}(2014)\citenamefont {Struck},
  \citenamefont {Simonet},\ and\ \citenamefont {Sengstock}}]{Struck}%
  \BibitemOpen
  \bibfield  {author} {\bibinfo {author} {\bibfnamefont {J.}~\bibnamefont
  {Struck}}, \bibinfo {author} {\bibfnamefont {J.}~\bibnamefont {Simonet}},\
  and\ \bibinfo {author} {\bibfnamefont {K.}~\bibnamefont {Sengstock}},\
  }\bibfield  {title} {\bibinfo {title} {Spin-orbit coupling in periodically
  driven optical lattices},\ }\href
  {https://doi.org/10.1103/PhysRevA.90.031601} {\bibfield  {journal} {\bibinfo
  {journal} {Phys. Rev. A}\ }\textbf {\bibinfo {volume} {90}},\ \bibinfo
  {pages} {031601} (\bibinfo {year} {2014})}\BibitemShut {NoStop}%
\bibitem [{\citenamefont {Goldman}\ and\ \citenamefont
  {Dalibard}(2014)}]{GoldmanPRX}%
  \BibitemOpen
  \bibfield  {author} {\bibinfo {author} {\bibfnamefont {N.}~\bibnamefont
  {Goldman}}\ and\ \bibinfo {author} {\bibfnamefont {J.}~\bibnamefont
  {Dalibard}},\ }\bibfield  {title} {\bibinfo {title} {Periodically {D}riven
  {Q}uantum {S}ystems: {E}ffective {H}amiltonians and {E}ngineered {G}auge
  {F}ields},\ }\href {https://doi.org/10.1103/PhysRevX.4.031027} {\bibfield
  {journal} {\bibinfo  {journal} {Phys. Rev. X}\ }\textbf {\bibinfo {volume}
  {4}},\ \bibinfo {pages} {031027} (\bibinfo {year} {2014})}\BibitemShut
  {NoStop}%
\bibitem [{\citenamefont {Fl\"aschner}\ \emph {et~al.}(2016)\citenamefont
  {Fl\"aschner}, \citenamefont {Rem}, \citenamefont {Tarnowski}, \citenamefont
  {Vogel}, \citenamefont {Lühmann}, \citenamefont {Sengstock},\ and\
  \citenamefont {Weitenberg}}]{Flaschner}%
  \BibitemOpen
  \bibfield  {author} {\bibinfo {author} {\bibfnamefont {N.}~\bibnamefont
  {Fl\"aschner}}, \bibinfo {author} {\bibfnamefont {B.~S.}\ \bibnamefont
  {Rem}}, \bibinfo {author} {\bibfnamefont {M.}~\bibnamefont {Tarnowski}},
  \bibinfo {author} {\bibfnamefont {D.}~\bibnamefont {Vogel}}, \bibinfo
  {author} {\bibfnamefont {D.-S.}\ \bibnamefont {Lühmann}}, \bibinfo {author}
  {\bibfnamefont {K.}~\bibnamefont {Sengstock}},\ and\ \bibinfo {author}
  {\bibfnamefont {C.}~\bibnamefont {Weitenberg}},\ }\bibfield  {title}
  {\bibinfo {title} {Experimental reconstruction of the {B}erry curvature in a
  {F}loquet {B}loch band},\ }\href {https://doi.org/10.1126/science.aad4568}
  {\bibfield  {journal} {\bibinfo  {journal} {Science}\ }\textbf {\bibinfo
  {volume} {352}},\ \bibinfo {pages} {1091} (\bibinfo {year}
  {2016})}\BibitemShut {NoStop}%
\bibitem [{\citenamefont {Ha}\ \emph {et~al.}(2015)\citenamefont {Ha},
  \citenamefont {Clark}, \citenamefont {Parker}, \citenamefont {Anderson},\
  and\ \citenamefont {Chin}}]{Ha}%
  \BibitemOpen
  \bibfield  {author} {\bibinfo {author} {\bibfnamefont {L.-C.}\ \bibnamefont
  {Ha}}, \bibinfo {author} {\bibfnamefont {L.~W.}\ \bibnamefont {Clark}},
  \bibinfo {author} {\bibfnamefont {C.~V.}\ \bibnamefont {Parker}}, \bibinfo
  {author} {\bibfnamefont {B.~M.}\ \bibnamefont {Anderson}},\ and\ \bibinfo
  {author} {\bibfnamefont {C.}~\bibnamefont {Chin}},\ }\bibfield  {title}
  {\bibinfo {title} {Roton-{M}axon {E}xcitation {S}pectrum of {B}ose
  {C}ondensates in a {S}haken {O}ptical {L}attice},\ }\href
  {https://doi.org/10.1103/PhysRevLett.114.055301} {\bibfield  {journal}
  {\bibinfo  {journal} {Phys. Rev. Lett.}\ }\textbf {\bibinfo {volume} {114}},\
  \bibinfo {pages} {055301} (\bibinfo {year} {2015})}\BibitemShut {NoStop}%
\bibitem [{\citenamefont {Schweizer}\ \emph {et~al.}(2019)\citenamefont
  {Schweizer}, \citenamefont {Grusdt}, \citenamefont {Berngruber},
  \citenamefont {Barbiero}, \citenamefont {Demler}, \citenamefont {Goldman},
  \citenamefont {Bloch},\ and\ \citenamefont {Aidelsburger}}]{Schweizer}%
  \BibitemOpen
  \bibfield  {author} {\bibinfo {author} {\bibfnamefont {C.}~\bibnamefont
  {Schweizer}}, \bibinfo {author} {\bibfnamefont {F.}~\bibnamefont {Grusdt}},
  \bibinfo {author} {\bibfnamefont {M.}~\bibnamefont {Berngruber}}, \bibinfo
  {author} {\bibfnamefont {L.}~\bibnamefont {Barbiero}}, \bibinfo {author}
  {\bibfnamefont {E.}~\bibnamefont {Demler}}, \bibinfo {author} {\bibfnamefont
  {N.}~\bibnamefont {Goldman}}, \bibinfo {author} {\bibfnamefont
  {I.}~\bibnamefont {Bloch}},\ and\ \bibinfo {author} {\bibfnamefont
  {M.}~\bibnamefont {Aidelsburger}},\ }\bibfield  {title} {\bibinfo {title}
  {Floquet approach to {Z}2 lattice gauge theories with ultracold atoms in
  optical lattices},\ }\href {https://doi.org/10.1038/s41567-019-0649-7}
  {\bibfield  {journal} {\bibinfo  {journal} {Nat. Phys.}\ }\textbf {\bibinfo
  {volume} {15}},\ \bibinfo {pages} {1168} (\bibinfo {year}
  {2019})}\BibitemShut {NoStop}%
\bibitem [{\citenamefont {Wintersperger}\ \emph {et~al.}(2020)\citenamefont
  {Wintersperger}, \citenamefont {Braun}, \citenamefont {\"Unal}, \citenamefont
  {Eckardt}, \citenamefont {Liberto}, \citenamefont {Goldman}, \citenamefont
  {Bloch},\ and\ \citenamefont {Aidelsburger}}]{Wintersperger}%
  \BibitemOpen
  \bibfield  {author} {\bibinfo {author} {\bibfnamefont {K.}~\bibnamefont
  {Wintersperger}}, \bibinfo {author} {\bibfnamefont {C.}~\bibnamefont
  {Braun}}, \bibinfo {author} {\bibfnamefont {F.~N.}\ \bibnamefont {\"Unal}},
  \bibinfo {author} {\bibfnamefont {A.}~\bibnamefont {Eckardt}}, \bibinfo
  {author} {\bibfnamefont {M.~D.}\ \bibnamefont {Liberto}}, \bibinfo {author}
  {\bibfnamefont {N.}~\bibnamefont {Goldman}}, \bibinfo {author} {\bibfnamefont
  {I.}~\bibnamefont {Bloch}},\ and\ \bibinfo {author} {\bibfnamefont
  {M.}~\bibnamefont {Aidelsburger}},\ }\bibfield  {title} {\bibinfo {title}
  {Realization of an anomalous {F}loquet topological system with ultracold
  atoms},\ }\href {https://doi.org/10.1038/s41567-020-0949-y} {\bibfield
  {journal} {\bibinfo  {journal} {Nat. Phys.}\ }\textbf {\bibinfo {volume}
  {16}},\ \bibinfo {pages} {1058} (\bibinfo {year} {2020})}\BibitemShut
  {NoStop}%
\bibitem [{\citenamefont {Mitchell}\ \emph {et~al.}(2021)\citenamefont
  {Mitchell}, \citenamefont {Di~Carli}, \citenamefont {Sinuco-Le\'on},
  \citenamefont {La~Rooij}, \citenamefont {Kuhr},\ and\ \citenamefont
  {Haller}}]{Mitchell}%
  \BibitemOpen
  \bibfield  {author} {\bibinfo {author} {\bibfnamefont {M.}~\bibnamefont
  {Mitchell}}, \bibinfo {author} {\bibfnamefont {A.}~\bibnamefont {Di~Carli}},
  \bibinfo {author} {\bibfnamefont {G.}~\bibnamefont {Sinuco-Le\'on}}, \bibinfo
  {author} {\bibfnamefont {A.}~\bibnamefont {La~Rooij}}, \bibinfo {author}
  {\bibfnamefont {S.}~\bibnamefont {Kuhr}},\ and\ \bibinfo {author}
  {\bibfnamefont {E.}~\bibnamefont {Haller}},\ }\bibfield  {title} {\bibinfo
  {title} {Floquet {S}olitons and {D}ynamics of {P}eriodically {D}riven
  {M}atter {W}aves with {N}egative {E}ffective {M}ass},\ }\href
  {https://doi.org/10.1103/PhysRevLett.127.243603} {\bibfield  {journal}
  {\bibinfo  {journal} {Phys. Rev. Lett.}\ }\textbf {\bibinfo {volume} {127}},\
  \bibinfo {pages} {243603} (\bibinfo {year} {2021})}\BibitemShut {NoStop}%
\bibitem [{\citenamefont {Lu}\ \emph {et~al.}(2022)\citenamefont {Lu},
  \citenamefont {Reid}, \citenamefont {Fritsch}, \citenamefont {Pi\~neiro},\
  and\ \citenamefont {Spielman}}]{LuM}%
  \BibitemOpen
  \bibfield  {author} {\bibinfo {author} {\bibfnamefont {M.}~\bibnamefont
  {Lu}}, \bibinfo {author} {\bibfnamefont {G.~H.}\ \bibnamefont {Reid}},
  \bibinfo {author} {\bibfnamefont {A.~R.}\ \bibnamefont {Fritsch}}, \bibinfo
  {author} {\bibfnamefont {A.~M.}\ \bibnamefont {Pi\~neiro}},\ and\ \bibinfo
  {author} {\bibfnamefont {I.~B.}\ \bibnamefont {Spielman}},\ }\bibfield
  {title} {\bibinfo {title} {Floquet {E}ngineering {T}opological {D}irac
  {B}ands},\ }\href {https://doi.org/10.1103/PhysRevLett.129.040402} {\bibfield
   {journal} {\bibinfo  {journal} {Phys. Rev. Lett.}\ }\textbf {\bibinfo
  {volume} {129}},\ \bibinfo {pages} {040402} (\bibinfo {year}
  {2022})}\BibitemShut {NoStop}%
\bibitem [{\citenamefont {Zhang}\ \emph {et~al.}(2013)\citenamefont {Zhang},
  \citenamefont {Chen},\ and\ \citenamefont {Zhang}}]{Zhang2013}%
  \BibitemOpen
  \bibfield  {author} {\bibinfo {author} {\bibfnamefont {Y.}~\bibnamefont
  {Zhang}}, \bibinfo {author} {\bibfnamefont {G.}~\bibnamefont {Chen}},\ and\
  \bibinfo {author} {\bibfnamefont {C.}~\bibnamefont {Zhang}},\ }\bibfield
  {title} {\bibinfo {title} {Tunable {S}pin-orbit {C}oupling and {Q}uantum
  {P}hase {T}ransition in a {T}rapped {B}ose-{E}instein {C}ondensate},\ }\href
  {https://doi.org/10.1038/srep01937} {\bibfield  {journal} {\bibinfo
  {journal} {Sci. Rep.}\ }\textbf {\bibinfo {volume} {3}},\ \bibinfo {pages}
  {1937} (\bibinfo {year} {2013})}\BibitemShut {NoStop}%
\bibitem [{\citenamefont {Jim\'enez-Garc\'{\i}a}\ \emph
  {et~al.}(2015)\citenamefont {Jim\'enez-Garc\'{\i}a}, \citenamefont {LeBlanc},
  \citenamefont {Williams}, \citenamefont {Beeler}, \citenamefont {Qu},
  \citenamefont {Gong}, \citenamefont {Zhang},\ and\ \citenamefont
  {Spielman}}]{Jimenez-Garcia}%
  \BibitemOpen
  \bibfield  {author} {\bibinfo {author} {\bibfnamefont {K.}~\bibnamefont
  {Jim\'enez-Garc\'{\i}a}}, \bibinfo {author} {\bibfnamefont {L.~J.}\
  \bibnamefont {LeBlanc}}, \bibinfo {author} {\bibfnamefont {R.~A.}\
  \bibnamefont {Williams}}, \bibinfo {author} {\bibfnamefont {M.~C.}\
  \bibnamefont {Beeler}}, \bibinfo {author} {\bibfnamefont {C.}~\bibnamefont
  {Qu}}, \bibinfo {author} {\bibfnamefont {M.}~\bibnamefont {Gong}}, \bibinfo
  {author} {\bibfnamefont {C.}~\bibnamefont {Zhang}},\ and\ \bibinfo {author}
  {\bibfnamefont {I.~B.}\ \bibnamefont {Spielman}},\ }\bibfield  {title}
  {\bibinfo {title} {Tunable {S}pin-{O}rbit {C}oupling via {S}trong {D}riving
  in {U}ltracold-{A}tom {S}ystems},\ }\href
  {https://doi.org/10.1103/PhysRevLett.114.125301} {\bibfield  {journal}
  {\bibinfo  {journal} {Phys. Rev. Lett.}\ }\textbf {\bibinfo {volume} {114}},\
  \bibinfo {pages} {125301} (\bibinfo {year} {2015})}\BibitemShut {NoStop}%
\bibitem [{\citenamefont {Gomez~Llorente}\ and\ \citenamefont
  {Plata}(2016)}]{Llorente}%
  \BibitemOpen
  \bibfield  {author} {\bibinfo {author} {\bibfnamefont {J.~M.}\ \bibnamefont
  {Gomez~Llorente}}\ and\ \bibinfo {author} {\bibfnamefont {J.}~\bibnamefont
  {Plata}},\ }\bibfield  {title} {\bibinfo {title} {Periodic driving control of
  {R}aman-induced spin-orbit coupling in {B}ose-{E}instein condensates: {T}he
  heating mechanisms},\ }\href {https://doi.org/10.1103/PhysRevA.93.063633}
  {\bibfield  {journal} {\bibinfo  {journal} {Phys. Rev. A}\ }\textbf {\bibinfo
  {volume} {93}},\ \bibinfo {pages} {063633} (\bibinfo {year}
  {2016})}\BibitemShut {NoStop}%
\bibitem [{\citenamefont {Huang}\ \emph {et~al.}(2018)\citenamefont {Huang},
  \citenamefont {Peng}, \citenamefont {Li}, \citenamefont {Meng}, \citenamefont
  {Chen}, \citenamefont {Qu}, \citenamefont {Wang}, \citenamefont {Zhang},\
  and\ \citenamefont {Zhang}}]{Huang2018}%
  \BibitemOpen
  \bibfield  {author} {\bibinfo {author} {\bibfnamefont {L.}~\bibnamefont
  {Huang}}, \bibinfo {author} {\bibfnamefont {P.}~\bibnamefont {Peng}},
  \bibinfo {author} {\bibfnamefont {D.}~\bibnamefont {Li}}, \bibinfo {author}
  {\bibfnamefont {Z.}~\bibnamefont {Meng}}, \bibinfo {author} {\bibfnamefont
  {L.}~\bibnamefont {Chen}}, \bibinfo {author} {\bibfnamefont {C.}~\bibnamefont
  {Qu}}, \bibinfo {author} {\bibfnamefont {P.}~\bibnamefont {Wang}}, \bibinfo
  {author} {\bibfnamefont {C.}~\bibnamefont {Zhang}},\ and\ \bibinfo {author}
  {\bibfnamefont {J.}~\bibnamefont {Zhang}},\ }\bibfield  {title} {\bibinfo
  {title} {Observation of {F}loquet bands in driven spin-orbit-coupled {F}ermi
  gases},\ }\href {https://doi.org/10.1103/PhysRevA.98.013615} {\bibfield
  {journal} {\bibinfo  {journal} {Phys. Rev. A}\ }\textbf {\bibinfo {volume}
  {98}},\ \bibinfo {pages} {013615} (\bibinfo {year} {2018})}\BibitemShut
  {NoStop}%
\bibitem [{\citenamefont {Zhang}\ \emph {et~al.}(2023)\citenamefont {Zhang},
  \citenamefont {Yi}, \citenamefont {Zhang}, \citenamefont {Jiao},
  \citenamefont {Shi}, \citenamefont {Yuan}, \citenamefont {Zhang},
  \citenamefont {Liu}, \citenamefont {Chen},\ and\ \citenamefont
  {Pan}}]{ZhangJY}%
  \BibitemOpen
  \bibfield  {author} {\bibinfo {author} {\bibfnamefont {J.-Y.}\ \bibnamefont
  {Zhang}}, \bibinfo {author} {\bibfnamefont {C.-R.}\ \bibnamefont {Yi}},
  \bibinfo {author} {\bibfnamefont {L.}~\bibnamefont {Zhang}}, \bibinfo
  {author} {\bibfnamefont {R.-H.}\ \bibnamefont {Jiao}}, \bibinfo {author}
  {\bibfnamefont {K.-Y.}\ \bibnamefont {Shi}}, \bibinfo {author} {\bibfnamefont
  {H.}~\bibnamefont {Yuan}}, \bibinfo {author} {\bibfnamefont {W.}~\bibnamefont
  {Zhang}}, \bibinfo {author} {\bibfnamefont {X.-J.}\ \bibnamefont {Liu}},
  \bibinfo {author} {\bibfnamefont {S.}~\bibnamefont {Chen}},\ and\ \bibinfo
  {author} {\bibfnamefont {J.-W.}\ \bibnamefont {Pan}},\ }\bibfield  {title}
  {\bibinfo {title} {Tuning {A}nomalous {F}loquet {T}opological {B}ands with
  {U}ltracold {A}toms},\ }\href
  {https://doi.org/10.1103/PhysRevLett.130.043201} {\bibfield  {journal}
  {\bibinfo  {journal} {Phys. Rev. Lett.}\ }\textbf {\bibinfo {volume} {130}},\
  \bibinfo {pages} {043201} (\bibinfo {year} {2023})}\BibitemShut {NoStop}%
\bibitem [{\citenamefont {Fujimoto}\ and\ \citenamefont
  {Uchino}(2019)}]{Kazuya}%
  \BibitemOpen
  \bibfield  {author} {\bibinfo {author} {\bibfnamefont {K.}~\bibnamefont
  {Fujimoto}}\ and\ \bibinfo {author} {\bibfnamefont {S.}~\bibnamefont
  {Uchino}},\ }\bibfield  {title} {\bibinfo {title} {Floquet spinor {B}ose
  gases},\ }\href {https://doi.org/10.1103/PhysRevResearch.1.033132} {\bibfield
   {journal} {\bibinfo  {journal} {Phys. Rev. Res.}\ }\textbf {\bibinfo
  {volume} {1}},\ \bibinfo {pages} {033132} (\bibinfo {year}
  {2019})}\BibitemShut {NoStop}%
\bibitem [{\citenamefont {Chen}\ \emph {et~al.}(2018)\citenamefont {Chen},
  \citenamefont {Wang}, \citenamefont {Li}, \citenamefont {Liu},\ and\
  \citenamefont {Hu}}]{Xiaolong}%
  \BibitemOpen
  \bibfield  {author} {\bibinfo {author} {\bibfnamefont {X.-L.}\ \bibnamefont
  {Chen}}, \bibinfo {author} {\bibfnamefont {J.}~\bibnamefont {Wang}}, \bibinfo
  {author} {\bibfnamefont {Y.}~\bibnamefont {Li}}, \bibinfo {author}
  {\bibfnamefont {X.-J.}\ \bibnamefont {Liu}},\ and\ \bibinfo {author}
  {\bibfnamefont {H.}~\bibnamefont {Hu}},\ }\bibfield  {title} {\bibinfo
  {title} {Quantum depletion and superfluid density of a supersolid in {R}aman
  spin-orbit-coupled {B}ose gases},\ }\href
  {https://doi.org/10.1103/PhysRevA.98.013614} {\bibfield  {journal} {\bibinfo
  {journal} {Phys. Rev. A}\ }\textbf {\bibinfo {volume} {98}},\ \bibinfo
  {pages} {013614} (\bibinfo {year} {2018})}\BibitemShut {NoStop}%
\bibitem [{\citenamefont {Lyu}\ and\ \citenamefont {Zhang}(2020)}]{Lyu}%
  \BibitemOpen
  \bibfield  {author} {\bibinfo {author} {\bibfnamefont {H.}~\bibnamefont
  {Lyu}}\ and\ \bibinfo {author} {\bibfnamefont {Y.}~\bibnamefont {Zhang}},\
  }\bibfield  {title} {\bibinfo {title} {Spin-orbit-coupling-assisted roton
  softening and superstripes in a {R}ydberg-dressed {B}ose-{E}instein
  condensate},\ }\href {https://doi.org/10.1103/PhysRevA.102.023327} {\bibfield
   {journal} {\bibinfo  {journal} {Phys. Rev. A}\ }\textbf {\bibinfo {volume}
  {102}},\ \bibinfo {pages} {023327} (\bibinfo {year} {2020})}\BibitemShut
  {NoStop}%
\bibitem [{\citenamefont {Li}\ \emph {et~al.}(2021)\citenamefont {Li},
  \citenamefont {Luo}, \citenamefont {Hou},\ and\ \citenamefont
  {Zhang}}]{Guanqiangli2021}%
  \BibitemOpen
  \bibfield  {author} {\bibinfo {author} {\bibfnamefont {G.-Q.}\ \bibnamefont
  {Li}}, \bibinfo {author} {\bibfnamefont {X.-W.}\ \bibnamefont {Luo}},
  \bibinfo {author} {\bibfnamefont {J.}~\bibnamefont {Hou}},\ and\ \bibinfo
  {author} {\bibfnamefont {C.}~\bibnamefont {Zhang}},\ }\bibfield  {title}
  {\bibinfo {title} {Pseudo-{G}oldstone excitations in a striped
  {B}ose-{E}instein condensate},\ }\href
  {https://doi.org/10.1103/PhysRevA.104.023311} {\bibfield  {journal} {\bibinfo
   {journal} {Phys. Rev. A}\ }\textbf {\bibinfo {volume} {104}},\ \bibinfo
  {pages} {023311} (\bibinfo {year} {2021})}\BibitemShut {NoStop}%
\end{thebibliography}%

\end{document}